\documentclass[aps,pre,twocolumn,floatfix]{revtex4-1}
\usepackage{graphicx}
\usepackage{placeins}
\usepackage{color}
\usepackage{latexsym}
\usepackage{amsmath}
\usepackage{amsfonts}
\usepackage{amssymb}
\usepackage{bm}
\usepackage{graphicx}
\usepackage{mathtools}
\usepackage{amsbsy}
\usepackage{physics}
\usepackage{enumitem}

\expandafter\ifx\csname package@font\endcsname\relax\else
 \expandafter\expandafter
 \expandafter\usepackage
 \expandafter\expandafter
 \expandafter{\csname package@font\endcsname}%
\fi
\usepackage{hyperref}

\usepackage[normalem]{ulem}

\graphicspath{{./figures/},{figures}}

\newcommand{\psum}{\sideset{}{'}\sum}

\let\Gamma\varGamma
\let\Delta\varDelta
\let\Theta\varTheta
\let\Lambda\varLambda
\let\Pi\varPi
\let\Sigma\varSigma
\let\Upsilon\varUpsilon
\let\Phi\varPhi
\let\Psi\varPsi
\let\Omega\varOmega

\begin{document}


\title[]{Optimal Transport of an Anisotropic Tracer in Dense Active Suspensions}

\author{Chandranshu Tiwari}
\email{chandranshu21@iiserb.ac.in}
\affiliation{Indian Institute of Science Education and Research, Bhopal, Madhya Pradesh, India}%

\author{Sunil P. Singh}
\email{spsingh@iiserb.ac.in}
\affiliation{Indian Institute of Science Education and Research, Bhopal, Madhya Pradesh, India}%

\date{\today}

\begin{abstract}
 The transport of anisotropic tracers in active fluids exhibits rich dynamical behavior arising from the interplay between particle shape, activity, and steric interactions. We employ Brownian dynamics simulations to investigate the motion of an elliptical tracer immersed in a suspension of active dumbbells. We find that both translational and rotational transport, characterized by the mean-square speed and diffusivity, are enhanced by more than an order of magnitude with increasing area fraction, $\phi$, of active dumbbells. Notably, tracer motion is enhanced along the major axis relative to the minor axis, with $\mathrm{v}_{\parallel}>\mathrm{v}_{\perp}$ and $D_{\parallel}>D_{\perp}$. Remarkably, both translational and rotational transport exhibit an optimum at an intermediate packing fraction of active dumbbells, with the corresponding transport coefficients decreasing at higher densities.
We show that this non-monotonic transport arises from the anisotropic accumulation and aggregation of active dumbbells around the tracer, which control the non-equilibrium force and torque fluctuations. Thus, establish a direct connection between the collective organization of active dumbbells at the tracer surface and its emergent translational and rotational transport.
 
\end{abstract}

 \maketitle
\section{Introduction}

The dispersion of a passive tracer in an active medium has attracted enormous attention, ranging from studying the equation of state of active systems to unraveling transport in complex environments. Such studies have significantly advanced our understanding of intracellular transport, non-equilibrium fluctuations generated by motor proteins,  bacterial suspensions, cytoplasmic flows, and related biological processes \cite{brangwynne2008cytoplasmic,dieterich2008anomalous,otten2012local,morozov2014enhanced,koslover2017cytoplasmic,souza2017anomalous,mogre2020getting,ziepke2022multi,lu2023go,feng2023unraveling,lin2025interactions}. Tracer particles also serve as a probe for chemical sensing in bacterial suspensions, nutrient absorption, and enhanced bio-mixing \cite{,kurtuldu2011enhancement,kapellos2022impact}. Beyond their role as passive probes of transport, tracers provide a unique means of characterizing the non-equilibrium fluctuations in active matter \cite{mallory2014curvature,kaiser2014transport,fodor2015activity,argun2016non,burkholder2017tracer,martin2018active,ye2020active,knevzevic2020effective,seyforth2022nonequilibrium,pellicciotta2025wall,daza2025diffusion,yadav2026behavior,paul2026interplay}. In particular, anisotropic tracers possess orientational degrees of freedom can couple to active stresses and collisions, thereby provide access to anisotropic diffusion and mechanical response those are absent in symmetric tracers.

The dynamical behavior of spherical tracers has been extensively studied in suspensions of microorganisms such as bacteria and algae \cite{wu2000particle,mino2011enhanced,drescher2011fluid,jepson2013enhanced,patteson2016particle,semeraro2018effective,ortlieb2019statistics,villalobos2025active}, active enzymes \cite{zhao2017enhanced}, as well as in synthetic active systems \cite{mino2011enhanced,caprini2024emergent}. They have revealed a variety of non-intuitive phenomena, including anomalous transport \cite{granek2022anomalous,codina2022small}, strong non-Gaussian statistics of displacements \cite{grossmann2024non,baule2023universal}, and violations of the fluctuation–dissipation and Stokes–Einstein relation \cite{chen2007fluctuations,maggi2017memory}. Furthermore, tracer diffusivities often exceed their thermal values by several orders of magnitude. Such enhancements may arise from a complex interplay of swimmer-induced hydrodynamic flows and direct steric collisions between swimmers and tracers \cite{nordanger2022anisotropic,kasyap2014hydrodynamic, underhill2008diffusion,mino2011enhanced,drescher2011fluid,jepson2013enhanced,patteson2016particle,lagarde2020colloidal,dhar2024active}.
Steric interactions alone can substantially enhance tracer transport and even be exploited to convert chemical energy into mechanical work in asymmetric gears in dense suspensions 
 \cite{angelani2009self,pietzonka2019autonomous}.

Most studies have focused on spherical tracers, whose dynamics are characterized solely by translational motion. However, many biological objects, colloidal probes, and synthetic micromachines possess anisotropic shapes. The broken rotational symmetry of such particles introduces an intrinsic coupling between translational and rotational degrees of freedom. Some recent experimental studies have investigated the dynamics of anisotropic tracers, including ellipsoids  \cite{peng2016diffusion,yang2016dynamics}, dumbbells \cite{von2021diffusive}, rods \cite{li2024anisotropic}, and chains \cite{aporvari2020anisotropic}, immersed in active fluids. Owing to their anisotropic shape, these tracers exhibit direction-dependent transport along different principal axes. Remarkably, experiments have reported that the diffusivity along the minor axis can exceed that along the major axis, in contrast to equilibrium dynamics \cite{peng2016diffusion}. This anomalous behavior has been attributed to the straining hydrodynamic flows generated by swimming bacteria in the far field. However, the underlying physical mechanism remains unclear. Numerical studies of point-like ellipsoidal tracers in microswimmer suspension suggest that far-field hydrodynamics alone may not fully account for the observed translational–rotational coupling and anisotropic diffusion \cite{nordanger2022anisotropic}. Consequently, it remains unclear whether the observed transport anomalies arise primarily from hydrodynamic interactions, steric collisions, or an interplay between the two.

Understanding the role of steric interactions particularly important because collisions between active agents and tracers are unavoidable in biological environments, concentrated active suspensions, and dense microbial communities. Isolating steric interactions provides a minimal framework for identifying the microscopic mechanisms underlying anisotropic transport, without the additional complexity introduced by far-field hydrodynamic interactions. Such an approach is therefore essential for disentangling collision-induced effects from hydrodynamic contributions and establishing a common description of tracer dynamics in active systems.  In particular, it is unclear how activity-induced surface accumulation, swimmer crowding, and tracer geometry combine to generate anisotropic forces and torques and how these microscopic interactions can determine translational and rotational transport properties. 

This work systematically investigates the transport properties of a passive elliptical tracer immersed in a suspension of active (self-propelled) dumbbells that interact exclusively via steric forces. Using Brownian dynamics simulations, we examine the effects of activity and concentration on the tracer's translational and rotational dynamics and elucidate how collisions and geometric anisotropy govern dynamics.  This represents one of the first systematic studies aimed at isolating the contribution of activity-induced collisions and surface accumulation to anisotropic tracer transport. We demonstrate that active dumbbells accumulate anisotropically around the tracer surface, generating activity-dependent forces and torques that strongly influence both translational and rotational motion. By systematically varying activity and concentration, we quantify the resulting transport properties and identify the role of surface accumulation and geometric anisotropy in producing direction-dependent diffusion. Our results provide a microscopic picture of how active collisions and crowding govern anisotropic tracer dynamics in active environments.

The article is organized as follows. In Section \ref{sec:model}, we describe the simulation model of the active dumbbells and the elliptical tracer, along with the relevant parameters. Section \ref{sec:results} presents the simulation results and is divided into three subsections: the aggregation of active dumbbells, the dynamics of the tracer in the laboratory frame, and the dynamics of the tracer in the body frame. Section \ref{sec:conclusion} discusses the implications of the results and summarizes the main findings. Details of the transformation of the tracer displacements and velocities from the laboratory frame to the body frame are provided in Appendix~\ref{app:frames}. The dynamics of a passive elliptical tracer are discussed in Appendix~\ref{app:passive}, while the analysis of the force variance acting on the tracer is presented in Appendix~\ref{app:diffusivity}.

\section{Simulation Model}\label{sec:model}
We consider an elliptical tracer in a bath of $N_d$ active dumbbells. Each active dumbbell comprises two identical monomers of diameter $\sigma$ connected by the harmonic potential
\begin{equation}
    U_s = \sum_{i=1}^{N_d}\frac{k_s}{2}(|\mathbf{r}_{2i} - \mathbf{r}_{2i-1}| - \ell_0)^2,
    \label{eq:1}
\end{equation}
where $\mathbf{r}_{2i}$ and $\mathbf{r}_{2i-1}$ are the position vectors  of monomers of the ${i}^{th}$ dumbbell ($i \in \{1,2,...N_d\}$), $\ell_0$ is the equilibrium bond length, and $k_s$ is the spring constant.  To avoid the overlap among various dumbbells, the excluded volume interaction is employed among dumbbell monomers by a truncated repulsive Lennard-Jones potential (LJ)
\begin{equation}
U_{LJ} = \sum_{i = 1}^{2N_d - 1}\sum_{j = i+1}^{2N_d} 4\epsilon \left[\left(\frac{\sigma}{r_{ij}}\right)^{12} - \left(\frac{\sigma}{r_{ij}}\right)^{6}  + \frac{1}{4} \right],
\label{Eq:LJ}
\end{equation}
where $r_{ij} = |\mathbf{r}_{i} - \mathbf{r}_{j}|$ is the distance between monomer pair $i$ and $j$, and $\epsilon$ is the strength of LJ repulsion. For a given pair of monomers, LJ potential is $U_{LJ}=0$ for $r_{ij} \ge  2^{1/6}\sigma$, else described by above Eq.~\ref{Eq:LJ}. 

The active force $f_a$ is imposed along the direction  $\hat{\mathbf{n}}_i$ of the bond vector connecting the two monomers of the dumbbell, where $\hat{\mathbf{n}}_i$ is the unit vector along the dumbbell axis, defined as $\hat{\mathbf{n}}_i = (\mathbf{r}_{2i} - \mathbf{r}_{2i-1})/(|\mathbf{r}_{2i} - \mathbf{r}_{2i-1}|)$ \cite{suma2014dynamics,suma2014motility}. 

The  equation of motion of a dumbbell monomer is given by the overdamped Langevin equation 
\begin{equation}
 \frac{\partial{\mathbf{r}_{i}}}{\partial{t}} = \frac{1}{{\gamma}_t} \left[- \mathbf{\nabla}_i  U_{LJ}  - \mathbf{\nabla}_{i}  U_s + \mathbf{F}_m^i + f_a \hat{\mathbf{n}}_i  \right] + \boldsymbol{\eta}_{i}^T,
 \label{Eq:motion}
\end{equation}
where  $\gamma_t$ is the viscous drag, ${\boldsymbol{\eta}_i}^T$ is Gaussian white noise having zero mean, and  its correlation is expressed in terms of the translational diffusion coefficient $D_T$, 
 $ \langle \eta_{\alpha i}^T(t) \cdot {{\eta}_{\beta j}}^T(t^{\prime}) \rangle = 2 D_T  \delta_{\alpha \beta} \delta_{i j} \delta(t-t^{\prime})$, with $\alpha, \beta \in \{x,y\}$,  and $\mathbf{F}_m^i$ is force acting on the monomer due to its interaction with the elliptical tracer which is short range in nature. 
 \begin{figure}[t]
    \centering
    \includegraphics[width = 0.95\columnwidth]{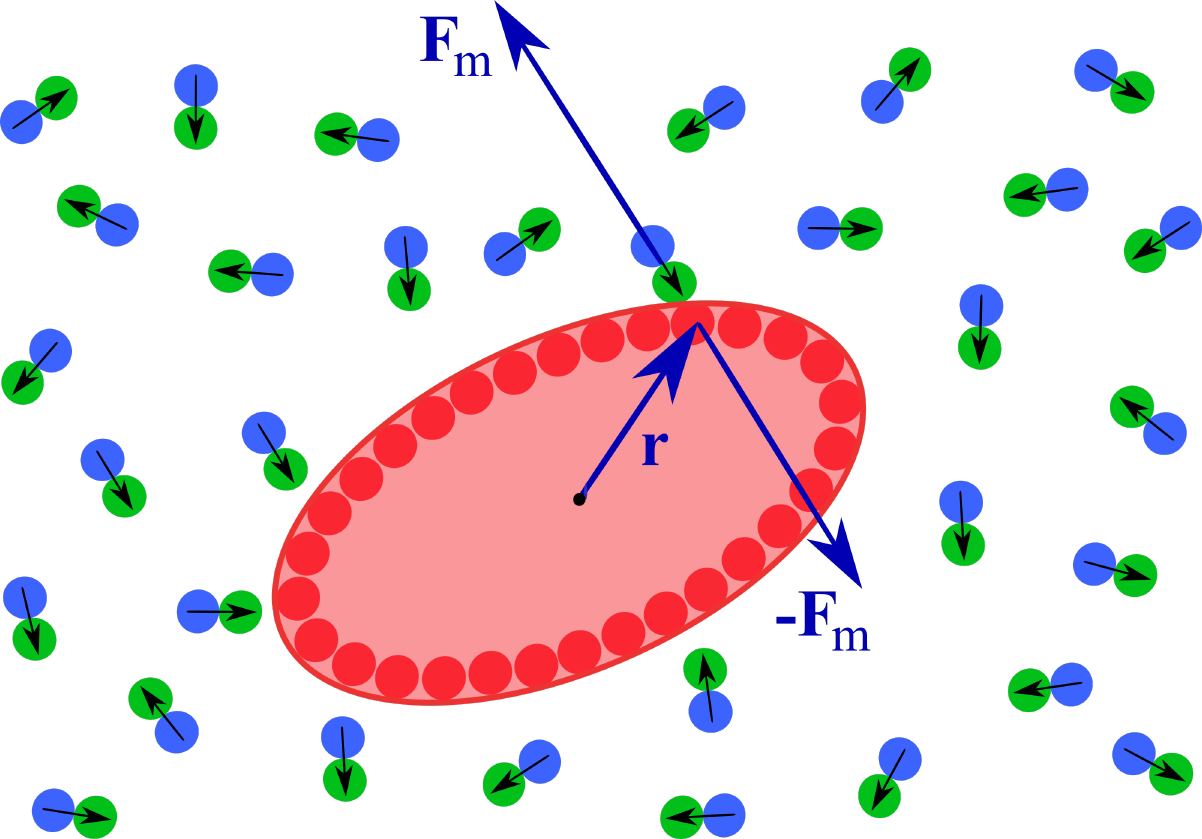}
    \caption{Schematic representation of the interaction between an active dumbbell and the tracer. The active dumbbell consists of green (head) and blue (tail) monomers. Red monomers present the hypothetical particles introduced on the tracer surface to model the tracer–dumbbell interaction. These particles appear in the simulation when a dumbbell monomer approaches the vicinity of the tracer.}
    \label{fig:schematic}
\end{figure}

The interaction between the dumbbell monomers and the tracer is also modeled via the short-range LJ potential given in Equation~\ref{Eq:LJ}. We employ a modified scheme to avoid the simulation artifact observed at higher P\'eclet numbers, particularly when the tracer particle is large, thereby effectively softening the tracer ellipsoid. To circumvent this issue, we effectively place hypothetical monomers identical to the actual monomers on the tracer surface. If the distance between a dumbbell monomer and the tracer surface is less than $(2^{1/6}(\sigma/2))$, the monomer interacts with these hypothetical monomers along the direction of its velocity, as illustrated in Figure~\ref{fig:schematic} by arrows with the force $\pm {\bf F}_m$.
The tracer experiences a net force due to these interactions, given by $\mathbf{F}_{e} = - \sum_{i=1}^{N_c}\mathbf{F}^i_m$, where $N_c$ is the number of monomers interacting with the tracer. Therefore, the equation of motion of the center of mass $\mathbf{R}_{cm}$ of the tracer is given by
 \begin{equation}
     M_{e}\frac{d^2 \mathbf{R}_{cm}}{d t^2} = \mathbf{F}_{e},
     \label{Eqn:trans_motion_tracer}
 \end{equation}
where $M_{e}$ is the mass of the tracer. Further, due to interaction forces, tracer also experiences a torque $\mathbf{\Gamma}_{e} = -\sum_{i=1}^{N_c}\mathbf{r}_i \times \mathbf{F}_{m}^i$ as shown in Figure~\ref{fig:schematic}.
The rotational motion of the tracer is governed by
 \begin{equation}
    I_{e} \frac{d^2 \theta }{d t^2} = \mathbf{\Gamma}_{e},
     \label{Eqn:orient_motion_tracer}
 \end{equation}
 where $I_e$ is the moment of inertia of the tracer, given as $I_e=M_e(a^2+b^2)/4$, with $a$ and $b$ representing the semi-major and semi-minor axes, respectively. The angle $\theta$ denotes the orientation of the semi-major axis with respect to the $x$-axis of the laboratory frame. The solvent-mediated long-range hydrodynamic interactions are neglected for the sake of simplicity of the model.

{\it Simulation Parameters:} All physical parameters are expressed in units of the LJ diameter $\sigma$, thermal energy $k_B T$, and the translational diffusion coefficient $D_T$. The simulation time is measured in units of $\tau = \sigma^2/D_T$. The mass of the tracer particle $M_e$ is expressed in units of dumbbell monomer mass $m$ as $M_e = (4 ab )m$ keeping the mass density fixed. The simulation parameters are chosen as follows: the simulation box size is $L/\sigma = 180$, the spring constant is $k_s = 20000 k_B T/\ell_0^2$, the dumbbell bond length is $\ell_0/\sigma = 1.0$, and the LJ interaction strength is $\epsilon/(k_B T) = 1 + Pe$. The Péclet number $Pe$, which characterizes the activity strength, is defined as the ratio of active to thermal force, $Pe = f_a \ell_0/(k_B T)$. The effective diffusivity of an active dumbbell in the dilute limit is $D_o^d = D^d_{cm} + f_a^2/(2\gamma^2D^d_R)$, where $D^d_{cm}$ and $D^d_R$ denote thermal center-of-mass and rotational diffusivities of the dumbbell, respectively \cite{suma2014dynamics}. The area fraction of dumbbells is defined as $\phi = (N_d \pi \sigma^2)/(8[L^2 - \pi ab])$, where $N_d$ denotes the number of dumbbells. The area fraction is varied in the range $0.07 \le \phi \le 0.5$. The aspect ratio of the elliptical tracer, $A_s = a/b$, is fixed to $1.5$ unless stated otherwise. For various packing fractions of the active dumbbells at a fixed box size, the number of dumbbells $N_d$ ranges from $1423$ to $10163$.  The  P\'eclet number $Pe$ is varied over the range $10$ to $25$.  The equations of motion for dumbbell monomers (Eq.~\ref{Eq:motion}) are integrated by the Euler algorithm with Gaussian distributed random displacements \cite{ermak1978brownian}, and equations of motion for tracer (Eq.~\ref{Eqn:trans_motion_tracer} and \ref{Eqn:orient_motion_tracer}) are integrated by the velocity-Verlet algorithm \cite{allen2017computer}. Integration steps ranging from $2.5 \times 10^{-5} \tau$ to $1 \times 10^{-5} \tau$. Each simulation point is averaged over at least ten independent simulation runs. Total simulation time ranges from $4000 \tau$ to $12000 \tau$.

 \begin{figure}[t]
    \centering
    \includegraphics[width = \columnwidth]{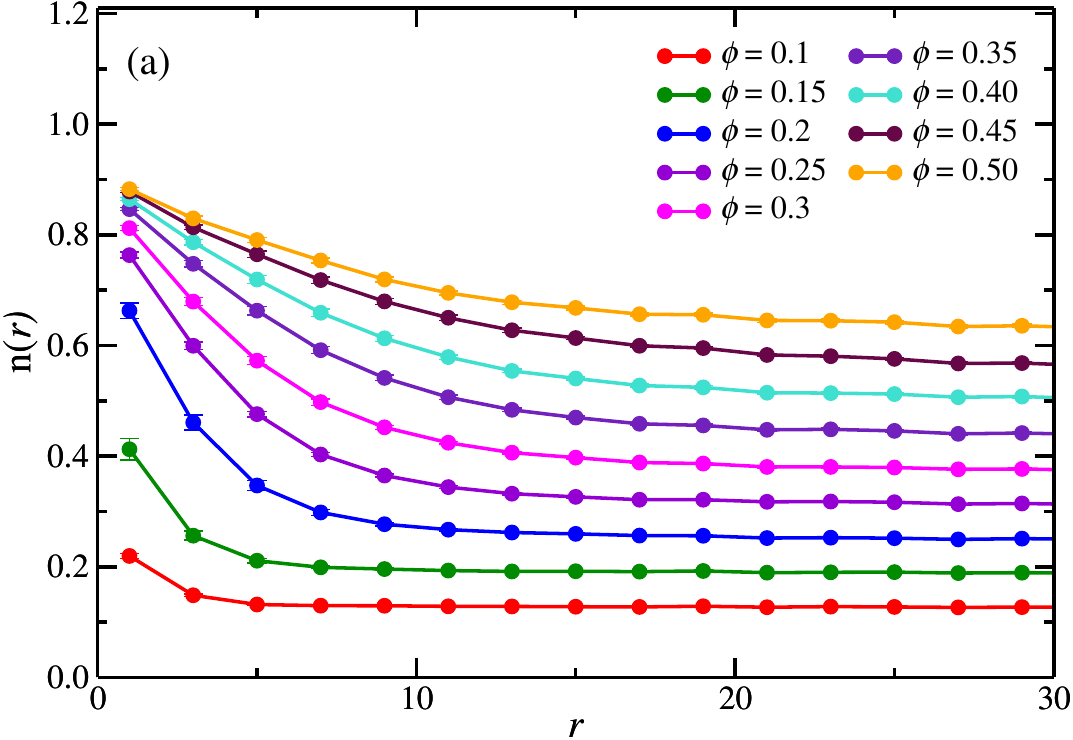}
    \includegraphics[width = \columnwidth]{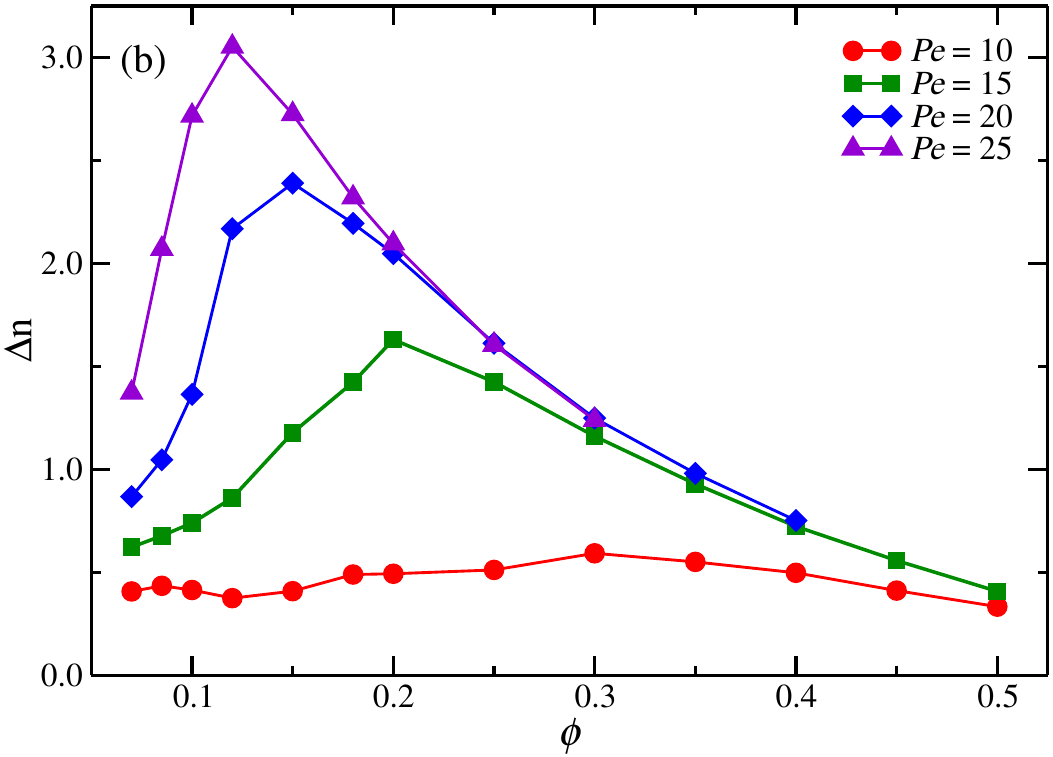} 
    \caption{(a) Density distribution of active dumbbells, $\mathrm{n}(r)$, around the elliptical tracer, measured as a function of distance from the surface of tracer for various area fractions $\phi$ at P\'eclet number $Pe=15$. (b) Relative excess density at the interface, $\Delta \mathrm{n} = (\mathrm{n}_1 - \mathrm{n}_0)/\mathrm{n}_0$ as a function of $\phi$ for various $Pe$. }
    \label{fig:rad_dist}
\end{figure}

\section{Results}\label{sec:results}
The dynamics of a passive anisotropic elliptical tracer immersed in a suspension of active dumbbells exhibits rich non-equilibrium behavior, including self-propelled motion and pronounced tumbling dynamics induced by the inhomogeneous accumulation of active dumbbells around the tracer. The supporting movies, Movie-S1 and Movie-S2, clearly illustrate these dynamical phenomena. To elucidate the underlying physical mechanisms governing the tracer dynamics, we first examine the aggregation behavior of active dumbbells on the surface of the elliptical tracer.



 \begin{figure*}[htp!]
    \centering
    \includegraphics[width = 0.318\textwidth]{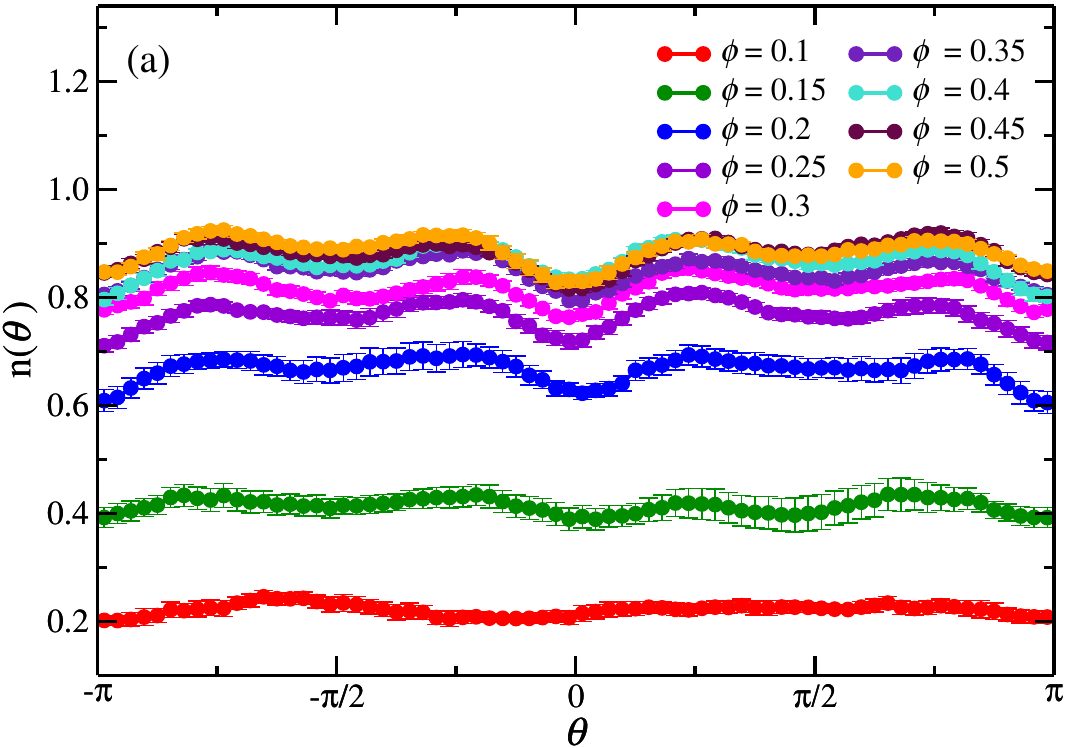}
    \includegraphics[width = 0.3175\textwidth]{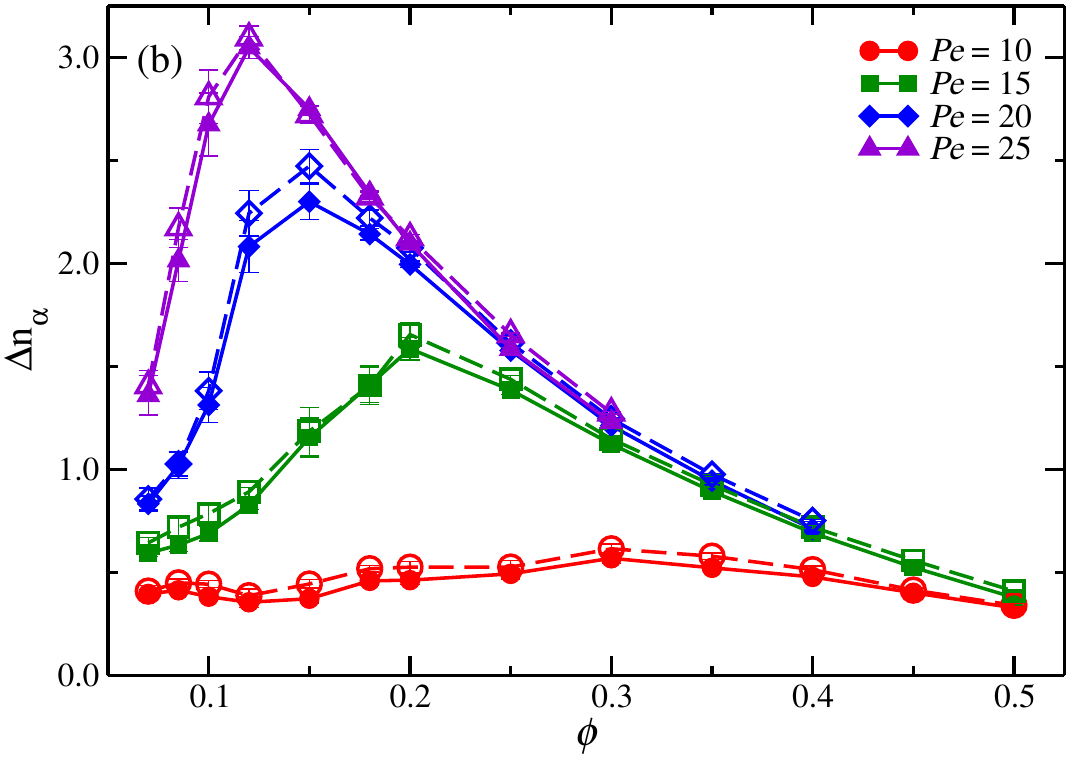}
    \includegraphics[width = 0.315\textwidth]{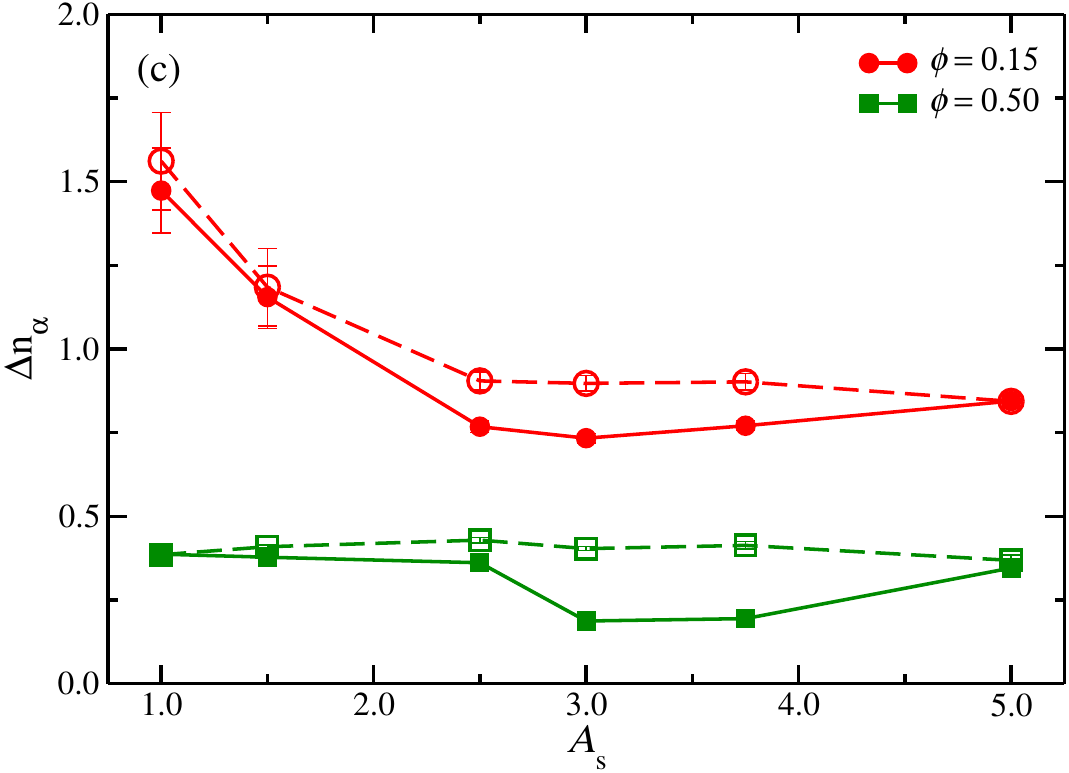}
    \caption{(a) Angular density profile of active dumbbells on the surface of tracer as a function of the polar angle $\theta$ measured from the semi-major axis,  for various area fractions $\phi$ at P\'eclet number  $Pe = 15$. (b) Variation of relative excess surface density of dumbbells $\Delta \mathrm{n}_\alpha = (\mathrm{n}_\alpha - \mathrm{n}_0)/\mathrm{n}_0$ as a function of $\phi$ for various $Pe$, where $\alpha \in \{\parallel, \perp\}$ represents the major and minor-axis directions, shown by solid and open symbols, respectively. For (a-b), the aspect ratio is fixed to $ A_s = 1.5$. (c) Variation of the excess surface density $\Delta \mathrm{n}_\alpha$ with aspect ratio $A_s$ at $Pe = 15$, with the semi-major axis fixed at $a = 15$. To demonstrate the influence of the active bath, we have considered both low and high packing fractions of the active dumbbells.}
    \label{fig:ang_dist}
\end{figure*}

\subsection{Aggregation on surface} 
We quantify the aggregation of active dumbbells around the elliptical tracer. Figure~\ref{fig:rad_dist}(a) presents the density distribution, $\mathrm{n}(r)$, as a function of the distance from the tracer surface, computed within concentric shells of thickness $\Delta r = 2$ (simulation units). The density profile shows a pronounced accumulation of active dumbbells near its surface, followed by a monotonic decay toward the bulk value for larger distances. 
The interfacial density, $\mathrm{n}_1 = \mathrm{n}(r = 1)$, increases systematically with the area fraction $\phi$, indicating enhanced aggregation of active dumbbells at the tracer surface. We expect this accumulation to become even more pronounced as activity increases.

The relative excess density at the interface as $\Delta \mathrm{n} = (\mathrm{n}_1-\mathrm{n}_0)/\mathrm{n}_0$, where $\mathrm{n}_0$ is the bulk number density is shown in   Figure~\ref{fig:rad_dist}(b).  The variation of the excess density $\Delta \mathrm{n}$  with $\phi$  exhibits a non-monotonic dependence.  It initially increases with area fraction, reaches a maximum at the crossover area fraction $\phi_c$, and then decreases. Furthermore, $\Delta \mathrm{n}$ systematically  increases  with the P\'eclet number  and for large $Pe$ and $\phi$, $\Delta \mathrm{n}$ saturates. This saturation appears to arise from the number of dumbbells at the tracer interface, reflecting the stronger accumulation of active dumbbells at higher activity levels. The crossover area fraction, $\phi_c$, beyond which $\Delta \mathrm{n}$ begins to decrease, shifts to lower values with increasing $Pe$, indicating that maximum interfacial accumulation is attained at lower packing fractions at higher activity.

The asymmetric shape of the tracer induces a heterogeneous angular distribution of active dumbbells along its surface. To examine this effect, we compute the angular density distribution of active dumbbells. The angular density profile is obtained by identifying dumbbells located within a distance of $\Delta r= 2$ (simulation units) from the surface into angular bins of $\Delta\theta =  5^\circ$, where the polar angle $\theta$ is measured from the semi-major axis of the tracer. Figure~\ref{fig:ang_dist}(a) presents the angular density profile as a function of $\theta$ for different area fractions $\phi$ of active dumbbells.  The profile exhibits a distinctly anisotropic pattern and density along the major axis ($\theta = 0$ and $\pi$) is significantly lower than the minor axis ($\theta = \pi/2$ and $-\pi/2$). The angular density profile becomes more heterogeneous and pronounced as the $\phi$ of active dumbbells in the suspension increases.

The heterogeneity of surface accumulation along the major and minor axes of the tracer can be further calculated from the excess surface densities $\Delta \mathrm{n}_{\alpha} = (\mathrm{n}_{\alpha} - \mathrm{n}_0)/\mathrm{n}_0$, where $\mathrm{n}_\alpha$ denotes the average surface density along the directions $\alpha \in \{\parallel,\perp\}$. Here $\parallel$ and $\perp$ represent the directions parallel and perpendicular to the semi-major axis, corresponding to the major and minor axes of the tracer, respectively.
The average densities are calculated as
\begin{align}\notag
\mathrm{n}_{\parallel}=
\frac{1}{\pi}\left[\sum_{-\pi/4}^{\pi/4} \mathrm{n}(\theta)\,\Delta\theta+\sum_{3\pi/4}^{5\pi/4} \mathrm{n}(\theta)\,\Delta\theta\right]; \\ 
\mathrm{n}_{\perp}=\frac{1}{\pi}\left[\sum_{\pi/4}^{3\pi/4}\mathrm{n}(\theta)\,\Delta\theta+\sum_{5\pi/4}^{7\pi/4} \mathrm{n}(\theta)\,\Delta\theta\right].
\end{align}
 Figure ~\ref{fig:ang_dist}(b) reveals that excess surface density on the minor-axis side remains consistently higher than that on the major-axis side. More importantly, the surface densities display non-monotonic variation with $\phi$ and the corresponding crossover area fraction, $\phi_c$, shifts to lower values for larger $Pe$, similar to the excess surface density $\Delta n$ (see Fig.~\ref{fig:rad_dist}(b)).

 \begin{figure}[ht]
    \centering
    \includegraphics[width = \columnwidth]{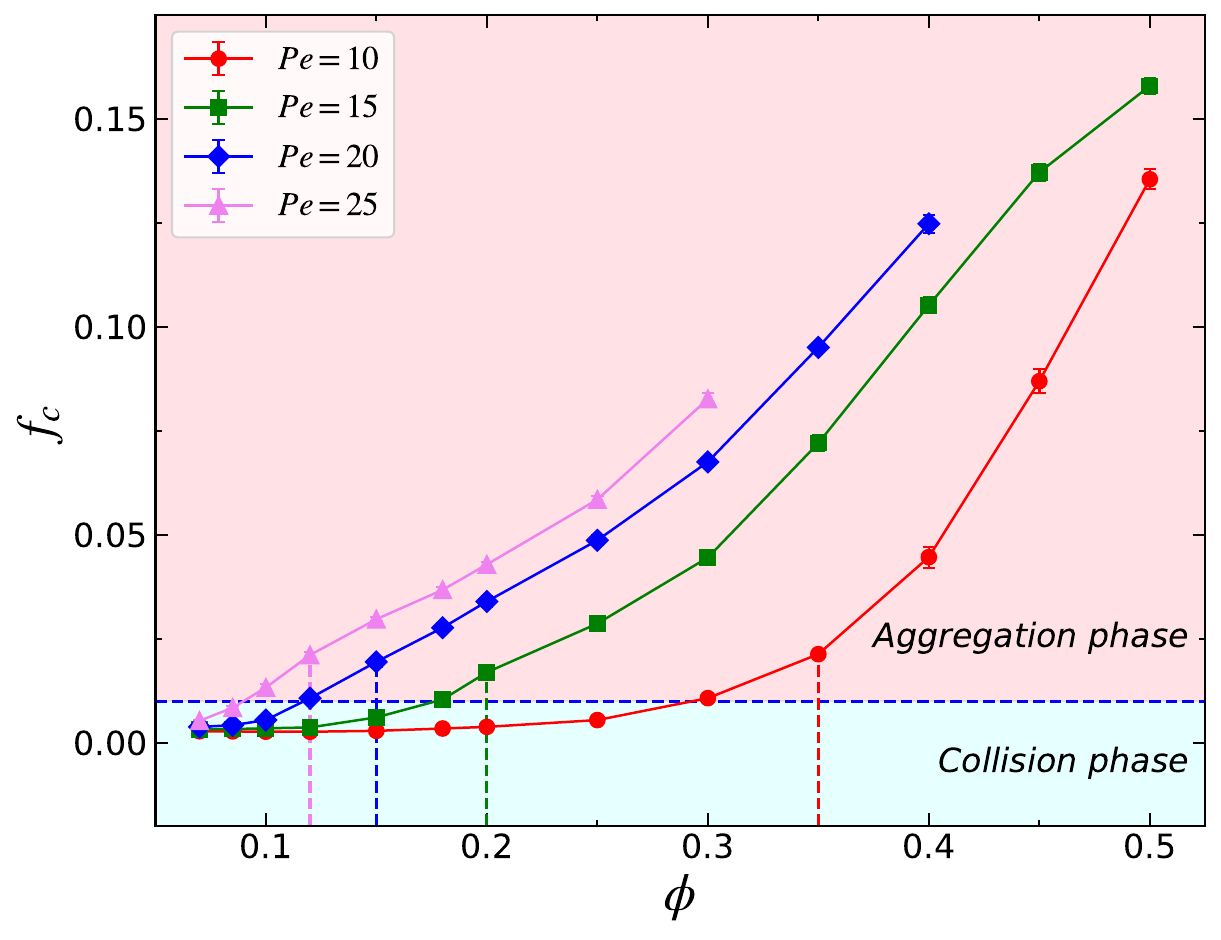}
    \caption{Average fraction of active dumbbells in the aggregate, $f_c = {\langle N_a\rangle}/{N_d}$ as a function of $\phi$ for various P\'eclet numbers ($Pe$). The blue dashed horizontal line indicates the threshold for each $Pe$, above which the aggregate size increases appreciably with increasing $\phi$; this region is highlighted in light pink. Vertical dashed lines show the critical area fractions $\phi_c$. Below this threshold, $f_c$ remains nearly zero, indicating the absence of aggregation, and is shown as the light cyan region. These two regimes correspond to the Aggregation phase and Collision phase, respectively.}
    \label{fig:aggregate}
\end{figure}

 Now we explore the role of the trace's aspect ratio $A_s$ on the aggregation behavior of the active dumbbells. Figure~\ref{fig:ang_dist}(c) displays the excess surface densities $\Delta n_\alpha$ as a function of $A_s$ while keeping the semi-major axis fixed at $a = 15\sigma$. The surface densities along the major and minor axes exhibit distinct profiles and show an overall decreasing trend with increasing aspect ratio $A_s$.   The difference between the densities along the major and minor axes increases with increasing $A_s$, indicating enhanced inhomogeneity. However, for sufficiently large aspect ratio ($A_s = 5$), the tracer becomes nearly rod-like $(a = 15, b = 3)$ and therefore the densities along the major and minor axes approach a homogeneous limit. 

 The overall aggregation of active dumbbells on the tracer surface is quantified using the following criterion: a dumbbell is considered to belong to an aggregate if at least one of its monomers is located within a distance of $1.2\sigma$ from a monomer of another dumbbell that is already part of the aggregate \cite{Spotswood_D_Stoddard,tiwari2024collective}. Figure \ref{fig:aggregate} shows the fraction of active dumbbells in the aggregate, $f_c = \langle N_a \rangle / N_d$, as a function of the area fraction $\phi$, where $N_a$ is the number of active dumbbells belonging to the aggregate. For a fixed activity $Pe$, $f_c$ is nearly zero at low $\phi$. However, beyond a critical area fraction $\phi_c$, $f_c$ increases rapidly, indicating the formation of larger aggregates on the tracer surface. Moreover, $f_c$ increases with increasing $Pe$ and the corresponding $\phi_c$ shifts to lower values, indicating that higher activity promotes aggregation through a mechanism reminiscent of motility-induced phase separation (MIPS) \cite{suma2014motility,cates2015motility,jose2021phase}. We refer to the regions above and below the critical area fraction as the aggregation phase and collision phase, respectively. Notably, for a given $Pe$, the critical area fraction for aggregation coincides with the crossover area fractions identified from the angular and radial density distributions in Figure~\ref{fig:ang_dist}(b) and \ref{fig:rad_dist}(b).

In summary, active dumbbells preferentially accumulate near the asymmetric tracer, particularly along its minor axis. The excess density increases up to a crossover area fraction, $\phi_c$, and decreases thereafter, with $\phi_c$ shifting to lower values as $Pe$ increases. This crossover coincides with the onset of stable aggregation. Increasing the tracer aspect ratio suppresses overall accumulation while enhancing spatial anisotropy.


\begin{figure*}[htp!]
    \centering
    \includegraphics[width = 0.32\textwidth]{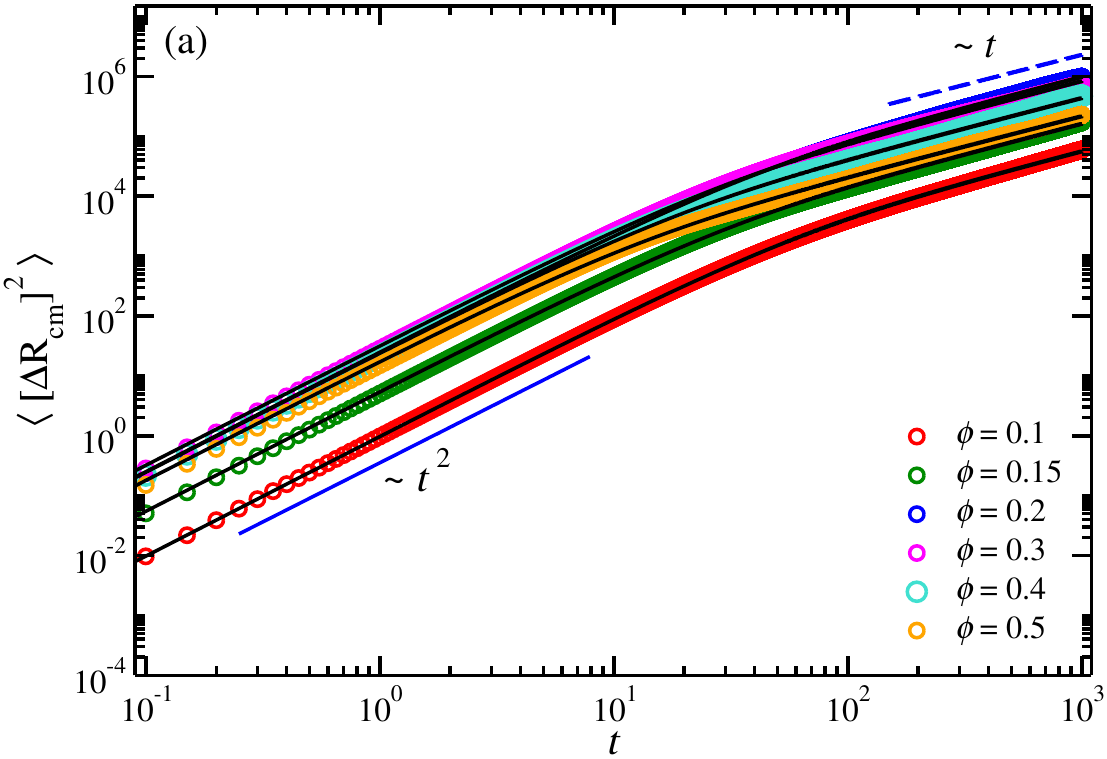}
    \includegraphics[width = 0.315\textwidth]{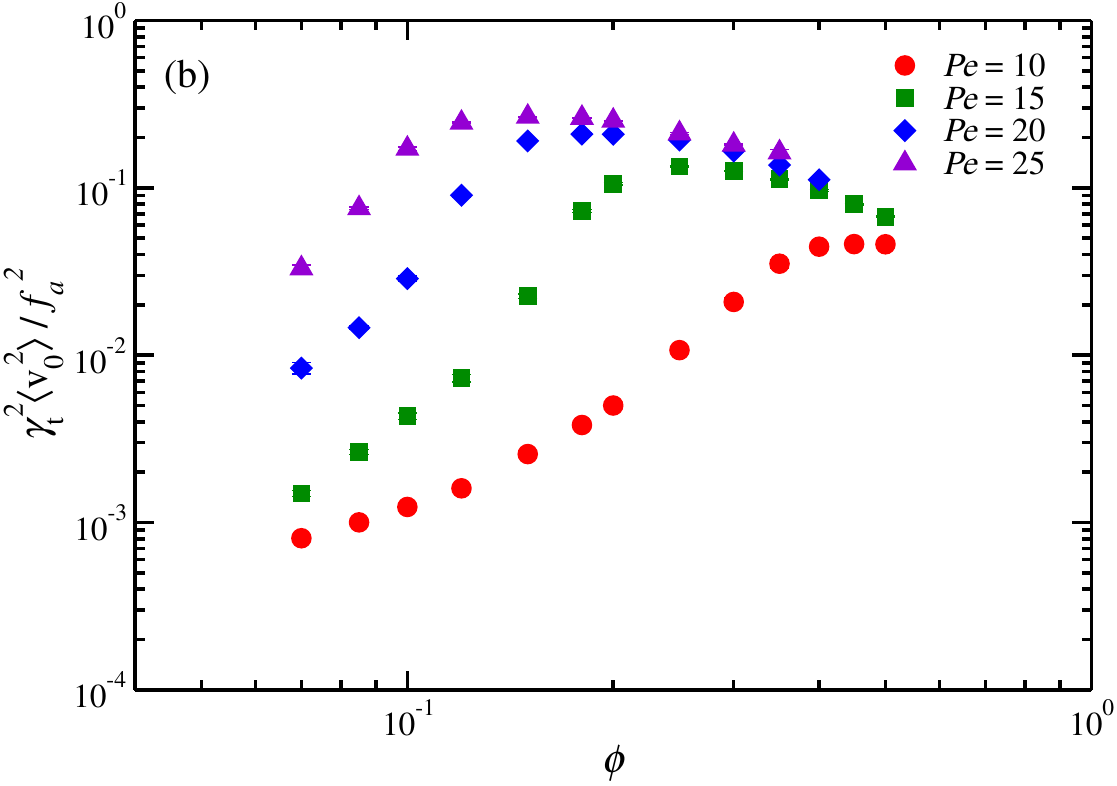}
    \includegraphics[width = 0.315\textwidth]{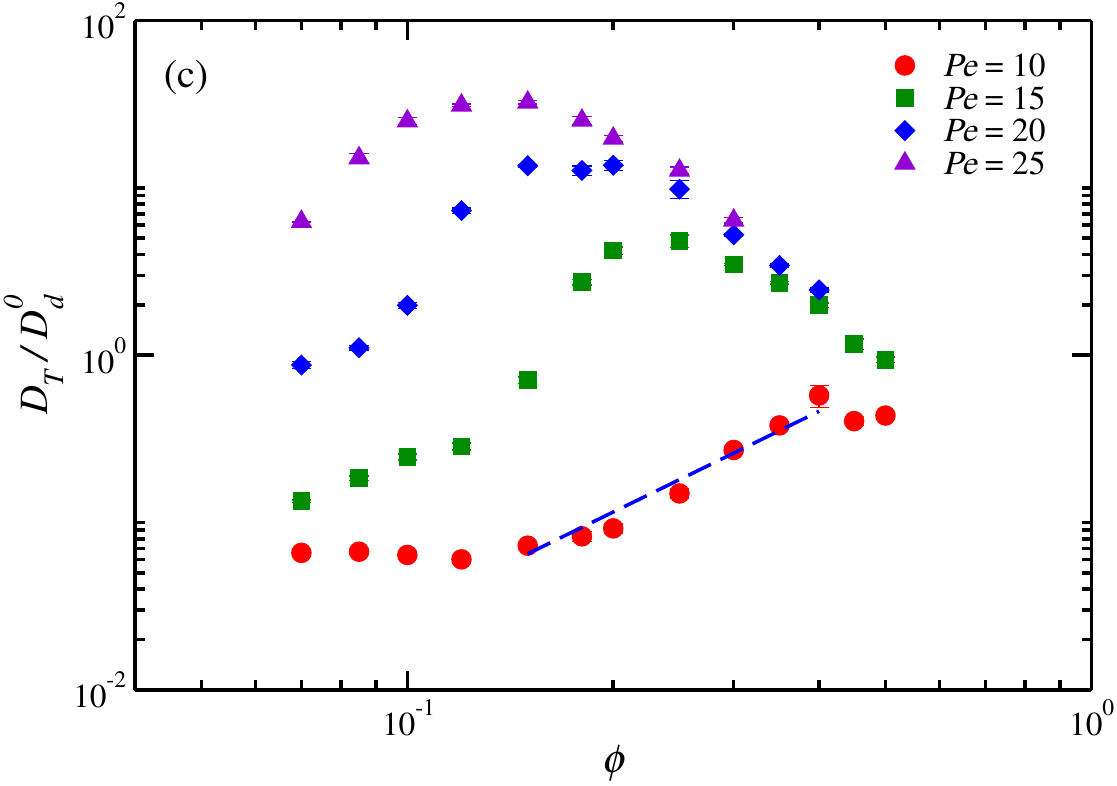}
    \caption{ (a) Mean-square displacement (MSD), $\langle[\Delta \mathbf{R}_{cm}]^2\rangle$, computed numerically (symbols) for various area fractions $\phi$ at a fixed P\'eclet number $Pe = 15$. The black lines are analytical fits of the MSD using Eq.~\ref{eq:OU_MSD}. The blue solid line indicates the ballistic behavior $\langle[\Delta \mathbf{R}_{cm}]^2\rangle\  \sim t^2$ and the blue dashed line shows the diffusive behavior $\langle[\Delta \mathbf{R}_{cm}]^2\rangle\  \sim t $ of the MSD.  (b) Non-monotonic variation of the scaled mean-square speed $\gamma_t^2\langle\mathrm{v_0}^2\rangle/f_a^2$ as a function of $\phi$ for various $Pe$, where $\gamma_t$ is the viscous drag and $f_a$ is the magnitude of the active force. (c) Non-monotonic behavior of the scaled translation diffusion coefficient $D_T/D^0_d$ with the area fraction of the active dumbbells ($\phi$) for various $Pe$. Here $D^0_d$ denotes the diffusivity of an active dumbbell in the dilute limit. The dashed line shows the quadratic dependence $D_T \sim \phi^2$.}
    \label{fig:msd}
\end{figure*}

\begin{figure}[t]
    \centering
    \includegraphics[width = \columnwidth]{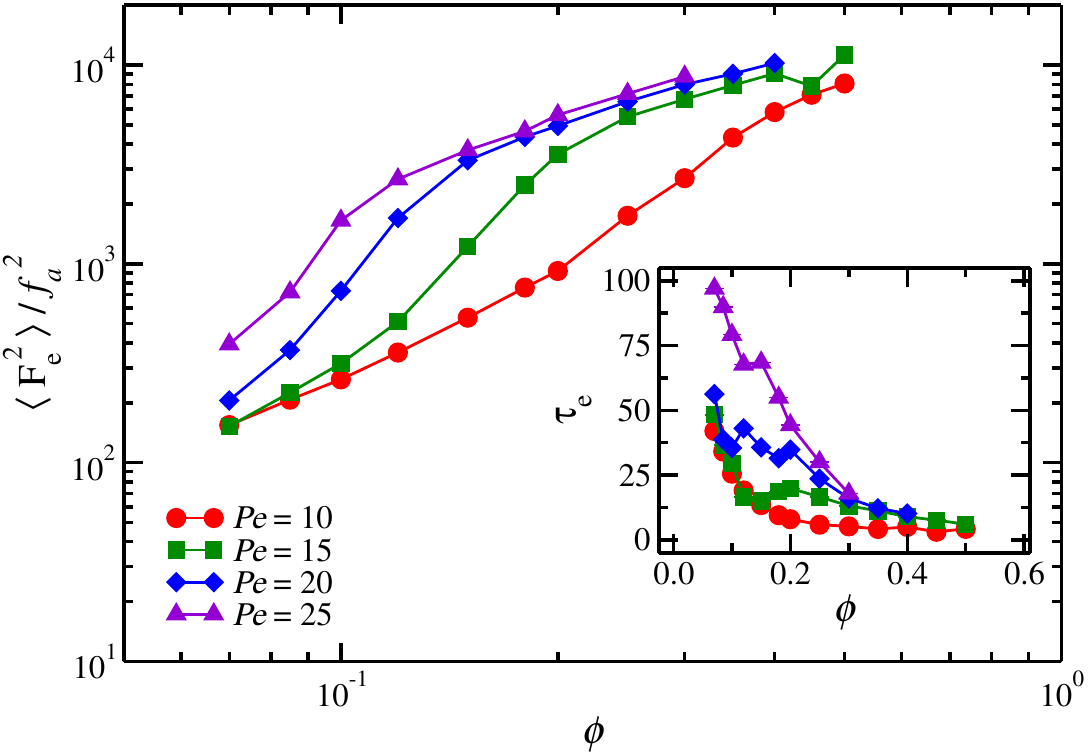}
    \caption{ The variance of the total force acting on the tracer as a function of $\phi$ for various $Pe$. The inset shows the crossover time $\tau_e$ in MSD  $\langle[\Delta \mathbf{R}_{cm}]^2\rangle$ from the transition from the ballistic to the diffusive regime.}
    \label{fig:cross_sq}
\end{figure}


The detailed dynamical behavior of the elliptic tracer, along with the microscopic origin of the observed behavior, is presented in the subsequent sections.

\subsection{Dynamics of tracer in laboratory frame}

The dynamics of the tracer emerge from the interplay between activity-induced collisions, surface accumulation, and the resulting force and torque fluctuations generated by active dumbbell suspension. In this section, we characterize the translational and rotational motion of the tracer in the laboratory frame.

\begin{figure*}[t]
    \centering
     \includegraphics[width = 0.32\textwidth]{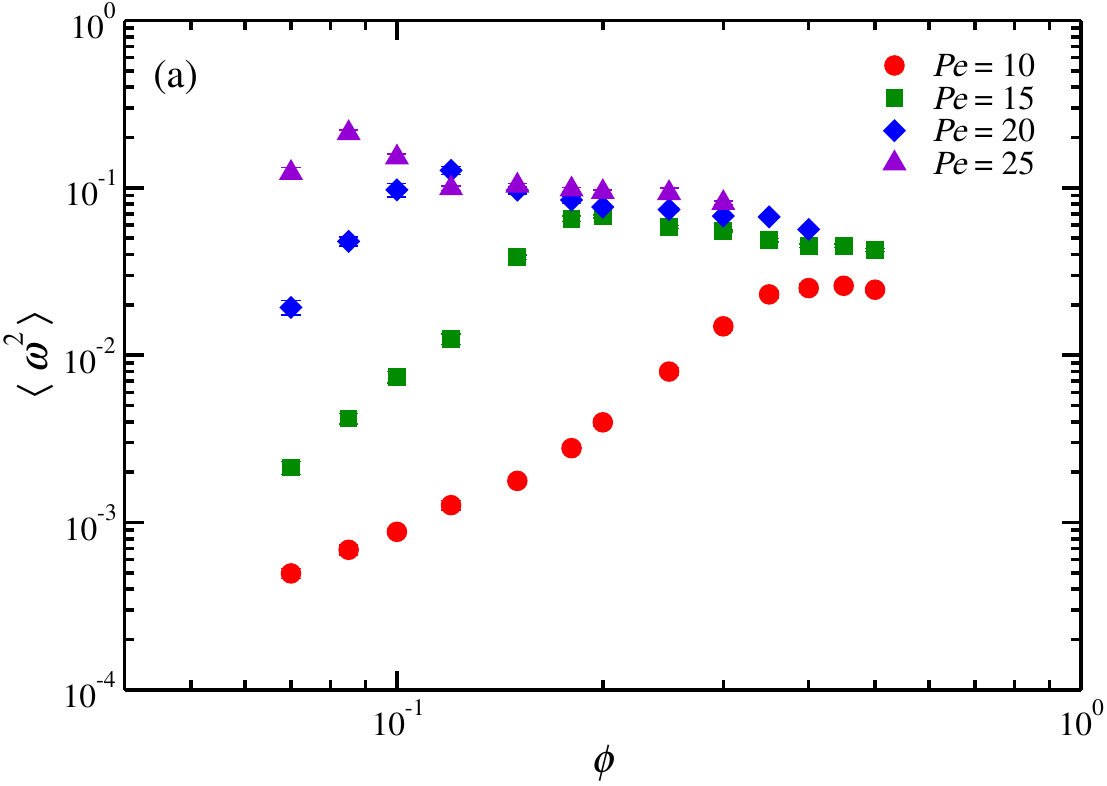}
    \includegraphics[width = 0.315\textwidth]{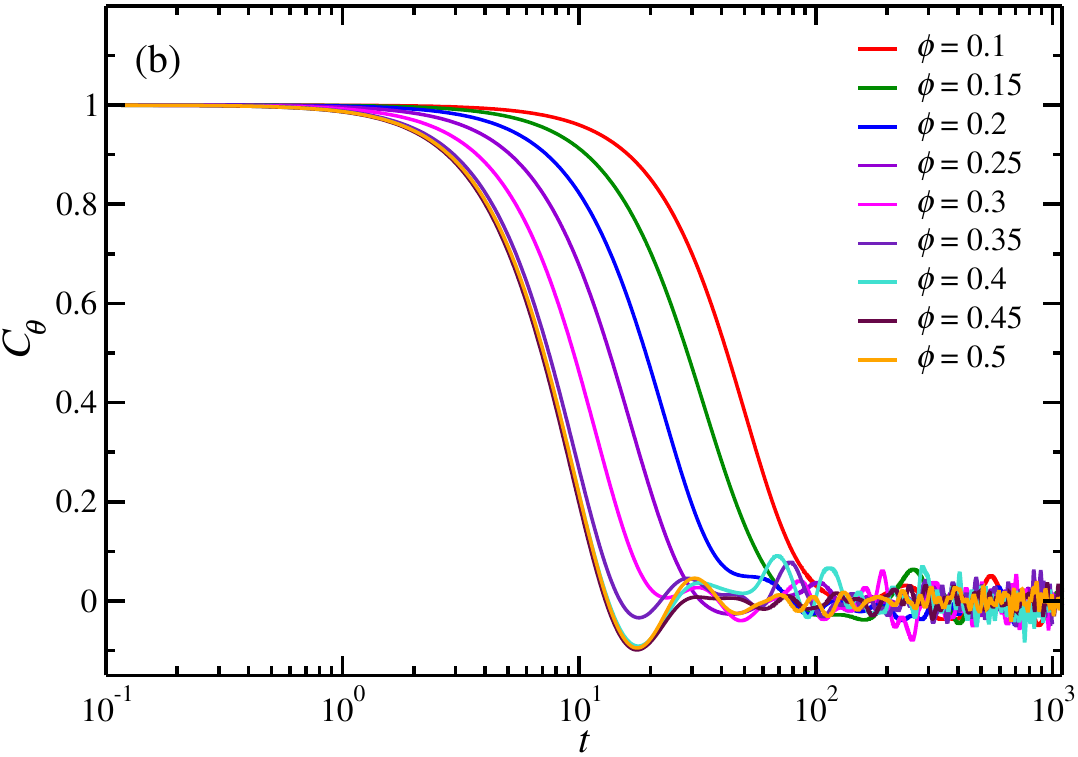}
    \includegraphics[width = 0.3175\textwidth]{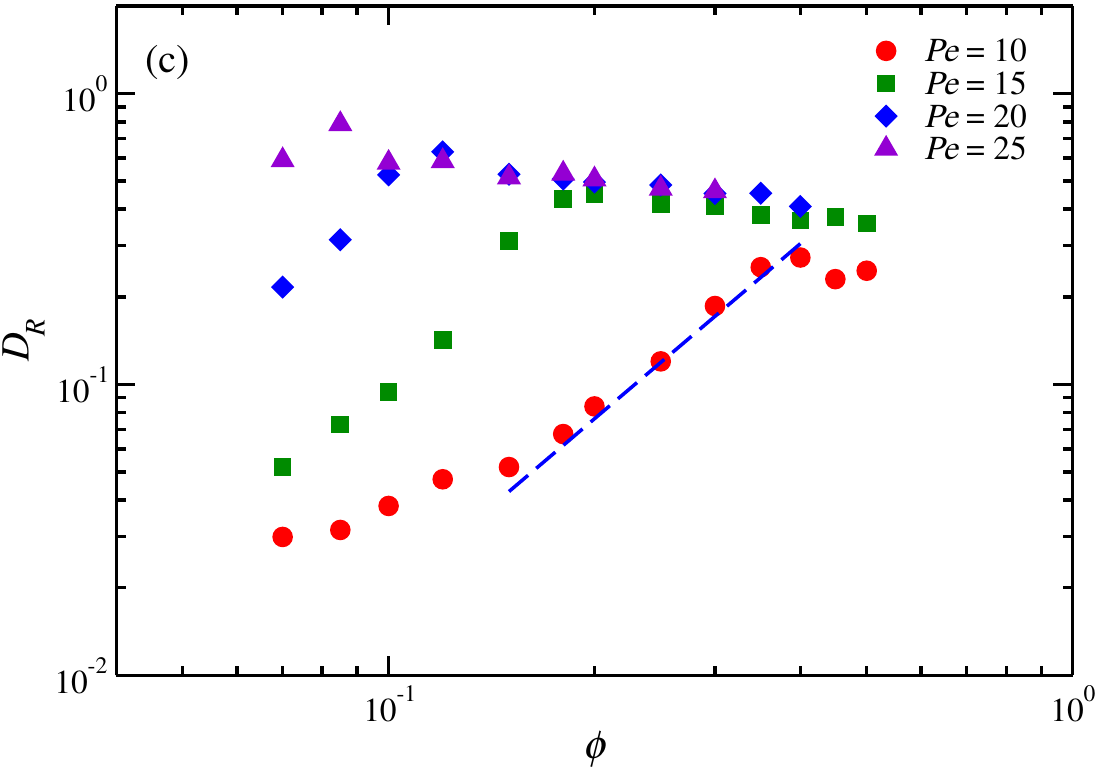}
    \caption{(a) Mean-square angular speed ($\langle\mathrm{\omega}^2\rangle$) as a function of area fraction $\phi$ for various P\'eclet numbers $(Pe)$. (b) Correlation of the semi-major axis of the tracer for various area fractions $\phi$ at P\'eclet number $Pe = 10$. (c) Rotational diffusion coefficient $D_R$ as a function of $\phi$ for various $Pe$; the dashed blue line shows the dependence $D_R \sim \phi^2$.  }
    \label{fig:theta_correl}
\end{figure*}

\subsubsection{Translational Dynamics}
To investigate the tracer's transport coefficients, we calculate the translational and rotational mean-square displacements (MSD). The translational MSD  is computed as
\begin{equation}
     \langle[\Delta \mathbf{R}_{cm}]^2\rangle =\ \langle[\mathbf{R}_{cm}(t+\Delta t) - \mathbf{R}_{cm}(t) ]^2\rangle,
\end{equation}
where $\mathbf{R}_{cm}$ is the position vector of the center of mass of the tracer.

Figure~\ref{fig:msd}(a) displays the MSD( $\langle[\Delta \mathbf{R}_{cm}]^2\rangle$)  of the tracer particle. At short times, the tracer exhibits ballistic motion, with $\langle[\Delta \mathbf{R}_{cm}]^2\rangle\  =\langle \mathrm{v}_0^2\rangle t^{2}$  where $\mathrm{v}_0$ is the speed of the tracer in the ballistic limit.
The MSD crosses over to the diffusive regime, $ \langle[\Delta \mathbf{R}_{cm}]^2\rangle\ \approx 4D_T t$, with $D_T$ as effective translational diffusivity,  qualitatively consistent with the experimental observations of passive tracer in the active medium \cite{peng2016diffusion, yang2016dynamics, li2024anisotropic}.

The mean-square speed $\langle \mathrm{v}_0^2\rangle$ of the tracer can be obtained from the ballistic regime by fitting $\langle[\Delta \mathbf{R}_{cm}]^2\rangle = \langle \mathrm{v}_0^2\rangle t^{2}$. Figure~\ref{fig:msd}(b) presents that  $\langle \mathrm{v}_0^2\rangle$ initially increases with the packing fraction $\phi$ of active dumbbells, reaches a maximum around a crossover area fraction $\phi_c$.  Beyond the crossover area fraction, $\langle \mathrm{v}_0^2\rangle$ decreases, while $\phi_c$ shifts to smaller values as activity increases. 

A similar non-monotonic dependence is observed in translational diffusivity, $D_T$, which is obtained by fitting $\langle[\Delta \mathbf{R}_{\mathrm{cm}}]^2\rangle = 4D_T t$, in the long-time diffusive regime of the MSD curves(see Fig.~\ref{fig:msd}(a)).  The $D_T$ initially increases at small $\phi$, reaches a maximum near the crossover area fraction $\phi_c$, and subsequently decreases beyond this point, as illustrated in Figure~\ref{fig:msd}(c).  The nonmonotonicity in diffusivity and speed becomes more pronounced at higher P'eclet numbers as displayed in Figure~\ref{fig:msd}(c).  
The increase in translational speed and diffusivity contrasts with passive systems (see Appendix Fig.\ref{fig:passive_results}(a-b)), where crowding of passive dumbbells always reduces tracer motion. 

The translational diffusivity displays a quadratic dependence  $D_T \sim \phi^2$, in the intermediate-density packing regime of active dumbbells,  consistent with experimental observations \cite{peng2016diffusion}. This quadratic increase of the effective diffusion is a signature of the giant-density fluctuations in active systems \cite{marchetti2013hydrodynamics}. Similar to the mean-square speed, the translational diffusivity increases with activity ($Pe$), and the crossover density $\phi_c$ shifts to lower values as $Pe$ increases. Notably, the effective diffusivity and the average speed of the passive tracer are enhanced by more than an order of magnitude in the active suspension, indicating that the enhanced transport is entirely driven by non-equilibrium active fluctuations of the medium.

To understand the physical mechanism underlying the non-monotonic behavior of the transport coefficient, we assume the effective force $\mathbf{F}_e$ exerted on the tracer by collisions with the surrounding active dumbbells as an active Ornstein--Uhlenbeck (OU) process \cite{villalobos2025active,martin2021statistical, szamel2014self,maggi2014generalized}

\begin{equation}
\tau_e\frac{d\mathbf{F}_e}{dt} = -\mathbf{F}_e + \gamma_e\boldsymbol{\Gamma}(t),
\label{eq:OU_force}
\end{equation}
where $\boldsymbol{\Gamma}(t)$ is Gaussian white noise with zero mean and correlations given as
\begin{equation}
\left\langle \Gamma_\alpha(t)\Gamma_\beta(t')\right\rangle = 2D_T\,\delta_{\alpha \beta}\,\delta(t-t'),
\end{equation}
with  $\alpha,\beta \in \{x, y\}$ and $D_T$ is effective diffusion coefficient. The stationary solution of Eq.~(\ref{eq:OU_force}) has zero mean and an exponentially decaying force correlations
\begin{equation}
\left\langle \mathrm{F}_{e,\alpha}(t) \mathrm{F}_{e,\beta}(t') \right\rangle = \delta_{\alpha \beta} \frac{\gamma_e^2D_T}{\tau_e} \exp\left( -\frac{|t-t'|}{\tau_e}\right).
\end{equation}

Here, $\tau_e$ is the timescale over which the force $\mathbf{F}_e$ decorrelates, and $\gamma_e$ is the effective drag on the tracer. Equivalently
\begin{equation}
    \left\langle\mathbf{F}_e(t)\cdot\mathbf{F}_e(t')\right\rangle = \langle \mathrm{F}_e^2\rangle \exp\left(-\frac{|t-t'|}{\tau_e}\right),
\label{eq:OU_corr}
\end{equation}
where $\langle F_e^2\rangle = (2 D_T\gamma_e^2 )/ \tau_e$ is total force variance in two dimension. 

The center-of-mass MSD of the tracer can be expressed in terms of the force autocorrelation \cite{zwanzig2001nonequilibrium} 
\begin{equation}
\langle[\Delta \mathbf{R}_{cm}]^2\rangle =  
\frac{1}{\gamma_e^2}\int^t_0 \int^t_0 \left\langle\mathbf{F}_e(s_1)\cdot\mathbf{F}_e(s_2)\right\rangle ds_1 ds_2
\label{eq:Green_kubo}
\end{equation}
which is further simplified as

\begin{equation}
\langle[\Delta \mathbf{R}_{cm}]^2\rangle = \frac{2\langle F_e^2\rangle\tau_e}{\gamma_e^2} \left[t -\tau_e\left(1-e^{-t/\tau_e}\right)\right].
\label{eq:OU_MSD}
\end{equation}

The expression for mean-square displacement (Eq. \ref{eq:OU_MSD}) agrees well with simulation results as presented in Figure~\ref{fig:msd} (a). The limiting forms of Equation \ref{eq:OU_MSD} recover the two regimes observed in the simulations. At short times ($t<\tau_e$), expanding the exponential gives $\left\langle \Delta R^2(t) \right\rangle \simeq (\langle F_e^2\rangle t^2)/\gamma_e^2$ indicating tracer moves ballistically, while at long times ($t>\tau_e$) the motion crosses over to diffusive regime, $\left\langle \Delta R^2(t) \right\rangle \simeq 4D_{T}t$, with effective diffusion coefficient

\begin{equation}
D_{\mathrm{T}} = \frac{\langle F_e^2\rangle\tau_e}{2\gamma_e^2}.
\end{equation}

The non-monotonic behavior of the effective diffusivity, $D_T$, can be understood in terms of the competition between the force variance, $\langle F_e^2\rangle$, and the force correlation time, $\tau_e$, as shown in Figure~\ref{fig:cross_sq}. At small $\phi$, the increase in $\langle F_e^2\rangle$ dominates over the reduction in the correlation time, $\tau_e$, resulting in an overall enhancement of the tracer speed and effective diffusivity.
However, as $\phi$ increases, $\langle F_e^2\rangle$ approaches a plateau due to the decrease in force cross-correlations at higher packing fractions. The individual self- and cross-correlation contributions are shown in Figure~\ref{fig:cross_sq_appendix}, revealing a pronounced decrease in the cross-correlation at large $\phi$, which causes $\langle F_e^2\rangle$ to saturate.  In contrast, $\tau_e$ continues to decrease. Consequently, the product $\langle F_e^2\rangle\tau_e$ decreases at large $\phi$, leading to a reduction in the tracer's diffusivity. Detailed discussion of the individual components of the force and cross-correlation is presented in Appendix C.


In summary, the non-monotonic behavior of effective diffusivity at high density arises from the saturation of force fluctuations and reduction in persistence time $\tau_e$ of random force exerted by the active dumbbells. In the next section, we focus on the aggregate's rotational behavior and its influence on the tracer's tumbling dynamics.

\subsubsection{Rotational Dynamics}
The asymmetric tracer exhibits fascinating dynamical behavior in the active medium, as illustrated in the supporting movies, Movies-S1 and Movie-S2. The heterogeneous density distribution of active dumbbells and their sliding motion, as seen in  Movie-S2, on the surface of the tracer induces an instantaneous torque, leading to spontaneous rotational motion. 
We analyze mean-square-angular-displacement (MSAD) to quantify the  rotational dynamics of the tracer
\begin{equation}
\langle[\Delta \theta]^2\rangle = \langle[\theta(t+\Delta t)-\theta(t)]^2\rangle,
\end{equation}
where $\theta$ is the orientation of the tracer’s semi-major axis with respect to the $x$-axis of the laboratory frame.

The MSAD, as expected,  exhibits a short-time ballistic regime followed by a long-time diffusive regime,  analogous to translational motion. The mean-square angular speed $\langle \omega^2\rangle$ of the tracer is obtained by fitting the expression in the ballistic regime $ \langle[\Delta \theta]^2\rangle \sim\langle \omega^2\rangle t^2$. The values shown in Figure~\ref{fig:theta_correl}(a) reveal once again a clear non-monotonic dependence on the area fraction $\phi$ of the active dumbbells: $\langle \omega^2\rangle$ increases at low densities, peaks near a crossover $\phi_c$, and decreases at higher $\phi$.   Increasing activity ($Pe$) amplifies the rotational fluctuations and shifts $\phi_c$ to lower values. A qualitatively similar non-monotonic dependence of the angular speed has been observed for microscopic asymmetric gears immersed in a bath of bacteria and Janus colloids \cite{sokolov2010swimming,maggi2016self}.

To complement this analysis, we further compute the correlation of the major-axis of the tracer
\begin{equation}
C_\theta (t) = \langle \cos[\theta(t) -\theta(0)] \rangle.
\end{equation}
Figure~\ref{fig:theta_correl}(b) illustrates $C_\theta(t)$ on a semi-logarithmic scale. For low packing fractions ($\phi \le 0.3$), the correlation decays exponentially as $C_\theta \sim \exp(-D_R t)$, where $D_R$ is the rotational diffusivity of the tracer. At larger $\phi$, the correlation develops damped oscillations for $t \ge 20$, with a few oscillatory periods that gradually diminish (see Fig.~\ref{fig:theta_correl}(b)).

Figure~\ref{fig:theta_correl}(c) shows that the rotational diffusivity $D_R$  increases sharply with the area fraction of active dumbbells $\phi$, reaches a maximum, and then decreases. Thus, $D_R$ displays a non-monotonic dependence on $\phi$, consistent with the behavior of the other rotational and translational measures reported in this work. The effective rotational diffusivity and the average angular speed of the tracer are enhanced by more than an order of magnitude in the active suspension, similar to the enhancements in translational diffusivity and speed. This infers that active fluctuations amplify not only translational transport but also the macroscopic rotational dynamics, thereby increasing the tracer's rotational diffusivity. At low activity ($Pe = 10$), $D_R$ varies quadratically with $\phi$ at intermediate densities, in agreement with experimental observations in active systems \cite{peng2016diffusion}. 

 The crossover behavior of both translational and rotational transport coefficients [Figs.~\ref{fig:msd}(b-c) and \ref{fig:theta_correl}(a,c)] coincides with the crossover in the aggregation of the active dumbbells (see Fig.~\ref{fig:aggregate}). At low area fractions $\phi$, active dumbbells interact with the tracer primarily through isolated collisions. Increasing $\phi$ enhances the frequency of these collisions, leading to larger force and torque fluctuations and, consequently, to higher translational and rotational transport coefficients. Beyond the crossover area fraction $\phi_c$, active dumbbells accumulate around the tracer surface (Fig.~\ref{fig:aggregate}) and hence constrain the tracer's translational and rotational motion. Therefore, activity-induced collisions, surface accumulation, and aggregation govern the translational and rotational transport of anisotropic tracers in dense active suspensions.

\begin{figure}[t]
    \centering
    \includegraphics[width = \columnwidth]{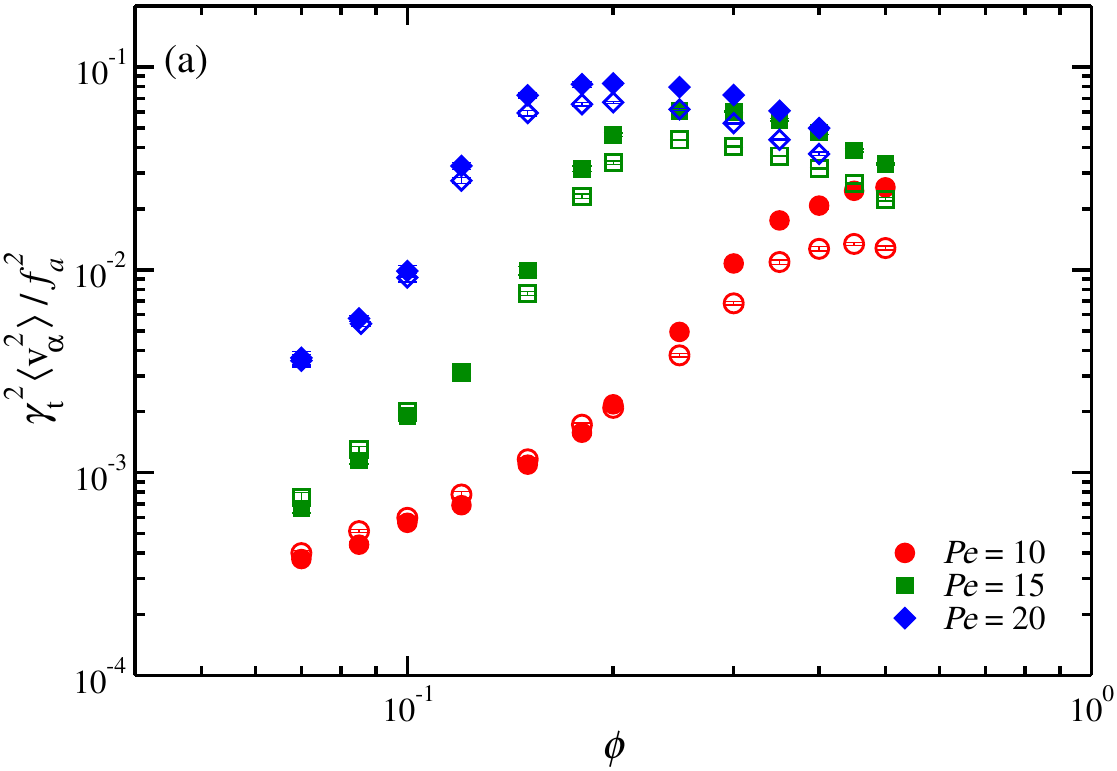}
    \includegraphics[width = \columnwidth]{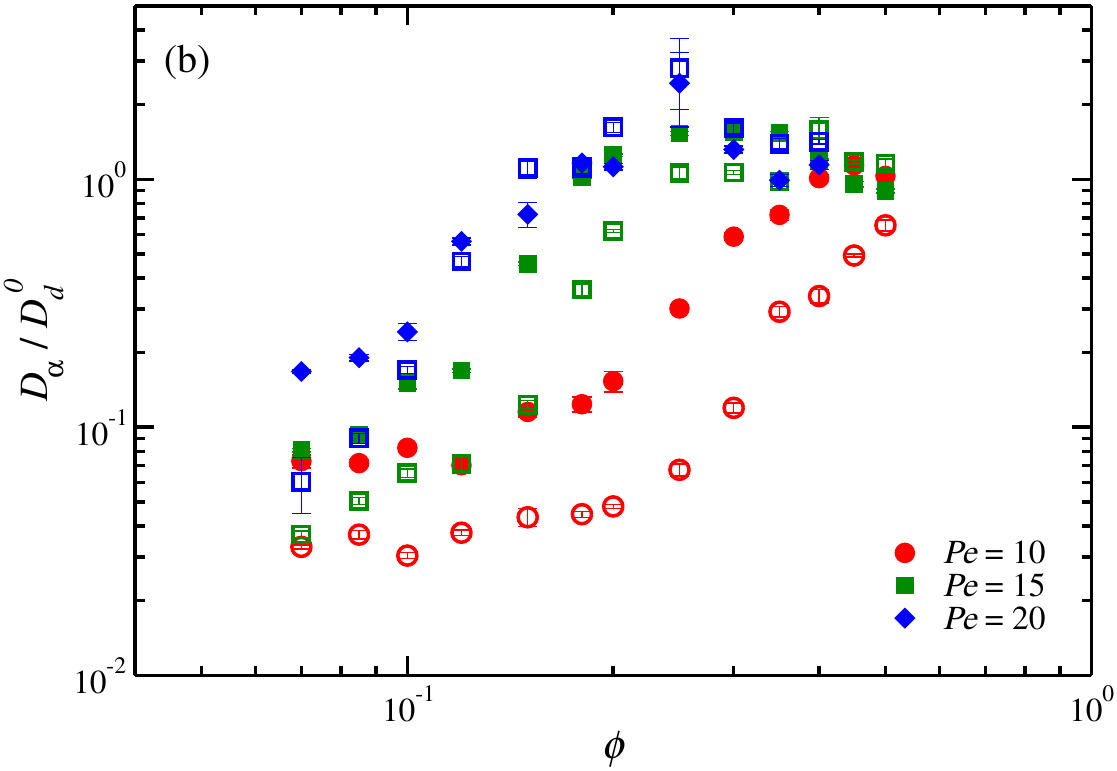}
    \caption{(a) Mean-square velocity and (b) diffusivity as functions of the area fraction $\phi$ for various activities $Pe$. $D^0_d$ is the diffusivity of an active dumbbell in the dilute limit. The components of the mean-square velocity and diffusivity parallel and perpendicular to the major axis, $\alpha \in \{\parallel,\perp\}$, are indicated by closed and open symbols, respectively.}
    \label{fig:msd_major}
\end{figure}

\begin{figure}[htp!]
    \centering
     \includegraphics[width = 0.975\columnwidth]{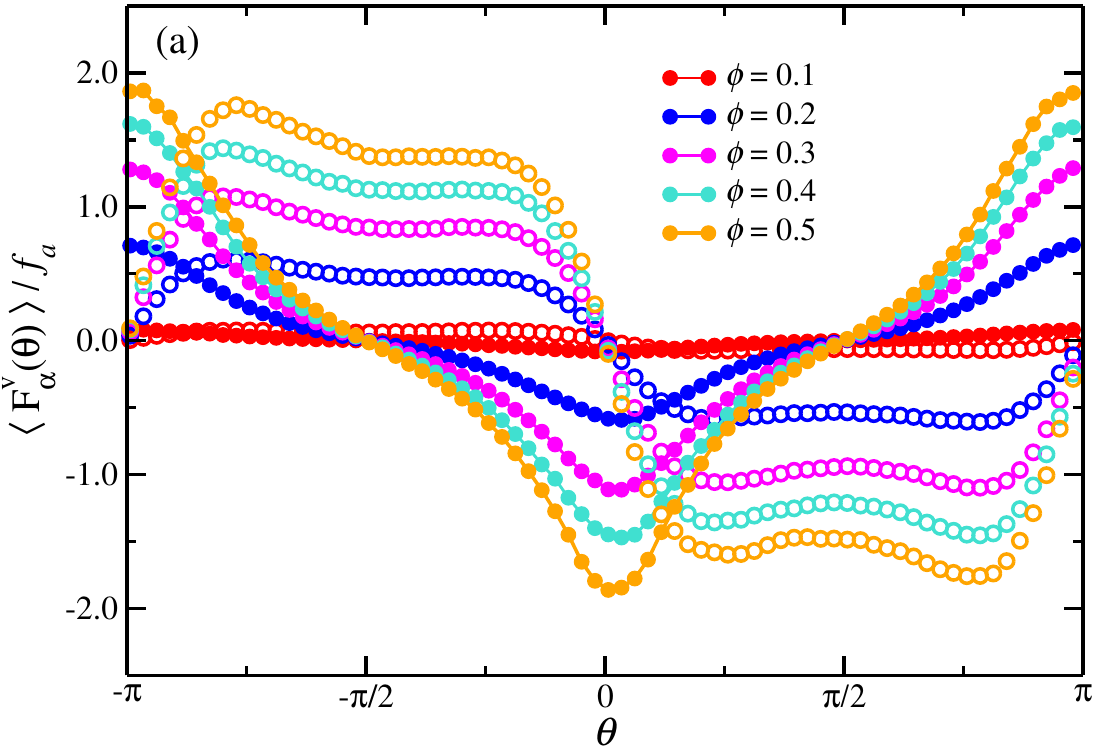}
     \includegraphics[width = 0.975\columnwidth]{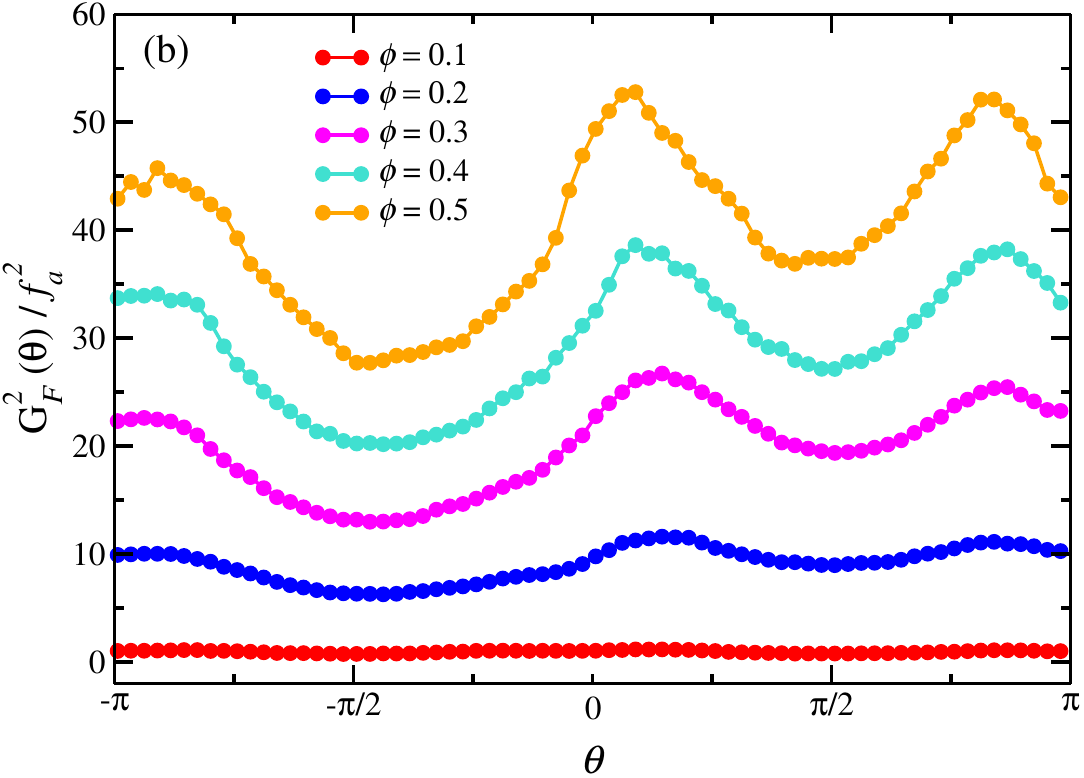}   
    \caption{
(a) Angular distribution of the scaled interaction force components $\langle \mathrm{F}^\mathrm{v}_\alpha(\theta)\rangle/f_a$ acting on the tracer for various area fractions $\phi$ at fixed $Pe = 15$. The angle $\theta$ is measured from the $\mathrm{v}_\parallel$ direction in the velocity--body frame, with $\alpha \in \{\parallel, \perp\}$ represented by solid and open symbols, respectively. (b) Angular distribution of the fluctuations of the total interaction force, defined as $\mathrm{G}_F^2(\theta) = \langle [\mathbf{F}^\mathrm{v}(\theta)]^2 \rangle - \langle \mathbf{F}^\mathrm{v}(\theta) \rangle^2$, where $\mathbf{F}^\mathrm{v}(\theta) = \mathrm{F}^\mathrm{v}_\parallel \mathbf{e}_{\mathrm{v}_\parallel} + \mathrm{F}^\mathrm{v}_\perp \mathbf{e}_{\mathrm{v}_\perp}$, for different values of $\phi$ at $Pe = 15$.
}
    \label{fig:force_dist}
\end{figure}

\subsection{Dynamics of tracer in the body frame}
The anisotropic shape of the tracer leads to distinct transport coefficients along its major and minor axes \cite{Doi2013SoftMP}. To quantify these features from the active agents' interactions, we analyze the tracer dynamics in the body-fixed (geometric) frame by transforming its motion from the laboratory frame. This transformation allows us to resolve the mean-square displacements (MSDs) along the tracer's principal axes. The details of the numerical procedure are provided in Appendix~\ref{app:frames}. 
Figure~\ref{fig:msd_major}(a) presents both components of the mean-square velocity  along the major and minor axes, obtained from the ballistic regime of the MSD according to $(\mathrm{MSD}\sim\langle \mathrm{v}_\alpha^2\rangle t^2)$, where $\alpha \in \{\parallel, \perp\}$. The corresponding long-time diffusivities, $D_\parallel$ and $D_\perp$, estimated from the diffusive regime of the MSD, are presented in Figure~\ref{fig:msd_major}(b). The mean-square velocities and diffusivities exhibit non-monotonic crossover with increasing area fraction $\phi$; they increase at low $\phi$, after approaching a maximum near the crossover density $\phi_c$, they decrease at larger $\phi$.

The mean-square velocity and diffusivity along the major axis remain larger than those along the minor axis over the explored range of parameters (Figs.~\ref{fig:msd_major}(a,b)). This behavior is consistent with equilibrium systems (Figs. \ref{fig:passive_results}(a,b)) and has also been reported for passive tracers in bacterial suspensions \cite{yang2016dynamics,nordanger2022anisotropic,aporvari2020anisotropic,li2024anisotropic}.

As shown in Figures ~\ref{fig:ang_dist}(a-c), the density of active dumbbells remains lower along the major axis compared to the minor axis. Consequently, the tracer experiences weaker steric interactions and reduced drag along its major axis, resulting in enhanced transport in this direction. This anisotropic transport is further reflected in the forces exerted by active dumbbells, which are resolved into the velocity body frame components $(\mathbf{e}_{\mathrm{v}_\parallel},\mathbf{e}_{\mathrm{v}_\perp})$, see appendix~\ref{app:frames}. Here, $\mathrm{F}^\mathrm{v}_\parallel$ and $\mathrm{F}^\mathrm{v}_\perp$ denote the forces exerted on the tracer by the active dumbbells along the $\mathbf{e}_{\mathrm{v}_\parallel}$ and $\mathbf{e}_{\mathrm{v}_\perp}$ directions, respectively. Figure~\ref{fig:force_dist}(a) illustrates that the average angular force distributions are strongly heterogeneous.  In particular, $\langle \mathrm{F}^\mathrm{v}_\parallel \rangle$ exhibits an antisymmetric angular dependence: it attains equal and opposite values at $\theta = 0$ and $\theta = \pm\pi$, and vanishes near $\theta = \pm\pi/2$. In contrast, $\langle \mathrm{F}^\mathrm{v}_\perp \rangle$ vanishes at $\theta = 0,\pm\pi$ and forms opposite-sign plateaus near $\pm\pi/2$.

The variance of the total force quantifies fluctuations and is directly related to anisotropic friction in the angular direction. Figure~\ref{fig:force_dist}(b) presents $\mathrm{G}_F^2(\theta) = \langle [\mathbf{F}^\mathrm{v}(\theta)]^2 \rangle - \langle \mathbf{F}^\mathrm{v}(\theta) \rangle^2$, which has peaks near $\theta = 0$ and $\pi$ and dips near $\pm\pi/2$, indicating enhanced fluctuations along the major axis. Therefore, $\mathrm{G}_F^2(\parallel)$  remains larger than $\mathrm{G}_F^2(\perp)$, resulting $\mathrm{v}_{\parallel} > \mathrm{v}_{\perp}$ and $\mathrm{D}_{\parallel} > \mathrm{D}_{\perp}$ across the explored parameter range. Increasing activity ($Pe$) further amplifies those force fluctuations, thereby enhancing both the translational and rotational motion of the passive tracer.

\section{Conclusion}\label{sec:conclusion}
We have systematically studied the dynamics of a passive elliptical tracer immersed in an active medium composed of active dumbbells. Our results demonstrate that the interplay between tracer anisotropy,  activity, and crowding governs emergent features of translational and rotational transport coefficients. Both translational and rotational motion exhibit a crossover from short-time ballistic motion to long-time diffusive behavior. The corresponding transport coefficients, namely the speed and diffusivity, are strongly enhanced in the active medium, increasing by nearly an order of magnitude with the density of active agents. Furthermore, we identify a non-monotonic dependence of the transport coefficients on the area fraction $\phi$, reflecting the competing effects of activity-enhanced collisions and crowding-induced hindrance. Our results show that the transport coefficients of the passive tracer are optimized at intermediate active-dumbbell densities, governed by the aggregation of active agents at the tracer surface and their correlations.

Activity drives the accumulation of dumbbells near the tracer surface, increasing the frequency of active collisions and thereby enhancing force and torque fluctuations. Consequently, both the instantaneous velocities and long-time diffusivities of the tracer increase at low to intermediate packing fractions. At higher packing fractions, however, the formation of stable surface aggregates leads to stronger anticorrelations between the forces exerted by different dumbbells, together with faster relaxation of the active forces. These effects suppress tracer mobility. The aggregation also reduces the anisotropic density imbalance, $\Delta n$, and constrains local motion, resulting in reduced translational and rotational velocities and diffusivities.


The key finding of this work is that the same physical mechanism governs both translational and rotational transport features. The anisotropic, activity-induced accumulation of dumbbells generates non-equilibrium forces and torques on the tracer, with the tracer geometry biasing these fluctuations. Because steric hindrance is weaker near the minor-axis region, the resulting force fluctuations preferentially drive motion along the tracer major axis, leading to $\mathrm{v}_{\parallel} > \mathrm{v}_{\perp}$ and $\mathrm{D}_{\parallel} > \mathrm{D}_{\perp}$ over most of the parameter range. Increasing the activity, characterized by the P\'eclet number $Pe$, amplifies these effects and shifts the crossover density $\phi_c$ to lower values.

Unlike previous studies that have primarily focused on dilute active baths \cite{wu2000particle,mino2011enhanced,drescher2011fluid,villalobos2025active,peng2016diffusion,yang2016dynamics,von2021diffusive,aporvari2020anisotropic}, our work highlights the crucial role of collective effects in dense active suspensions and the geometric anisotropy of both the bath and tracer particles. In particular, we demonstrate that activity-induced aggregation is not merely a structural feature but a central mechanism governing tracer transport. Our results directly connect microscale collective organization and aggregation with the emergent transport of anisotropic tracers in dense active systems. These findings broaden our understanding of transport in crowded active environments, including bacterial suspensions, active colloidal materials, and intracellular systems. Future studies incorporating long-range hydrodynamic interactions, external flows \cite{peng2016diffusion,li2024anisotropic,nordanger2022anisotropic}, and the effect of curvature \cite{nikola2016active,mallory2014curvature,pellicciotta2025wall} may reveal new dynamical behavior arising from the interplay between active stresses, tracer anisotropy, and confinement geometry.


\section*{SUPPLEMENTARY MATERIAL}
The supplementary material contains two supporting movies. Movie-S1 shows the dynamics of the elliptical tracer in the laboratory frame, while Movie-S2 illustrates the dynamics in the center-of-mass frame. The area fraction of active dumbbells is $\phi = 0.25$, and the activity is $Pe = 20$.

\begin{acknowledgments}

SPS acknowledges funding support from the DST-SERB Grant No. CRG/2020/000661. CHT acknowledges UGC India for the fellowship and the HPC facilities at IISER Bhopal and the Param Himalaya NSM facility. SPS acknowledges the hospitality of MPIPKS Dresden during his visit the draft of the manuscript was prepared.
\end{acknowledgments} 

\section*{AUTHOR DECLARATIONS}
The data are not publicly available. The data are available from the authors upon reasonable request.
\section*{Conflict of Interest}

The authors have no conflicts of interest to declare.

\section*{Author contributions}
 SPS designed and supervised the research project. CHT developed the simulation code and analyzed the simulation results. All authors contributed to the interpretation of the data, discussion of the results, and writing of the manuscript.

\appendix

\section{Transformation from laboratory to body frame.}
\label{app:frames}
The dynamics of the tracer can be analyzed using two distinct body frames: the geometric body frame and the velocity body frame.

\renewcommand{\thefigure}{A\arabic{figure}}
\setcounter{figure}{0}

\begin{figure}[htp]
    \centering
    \includegraphics[width = \columnwidth]{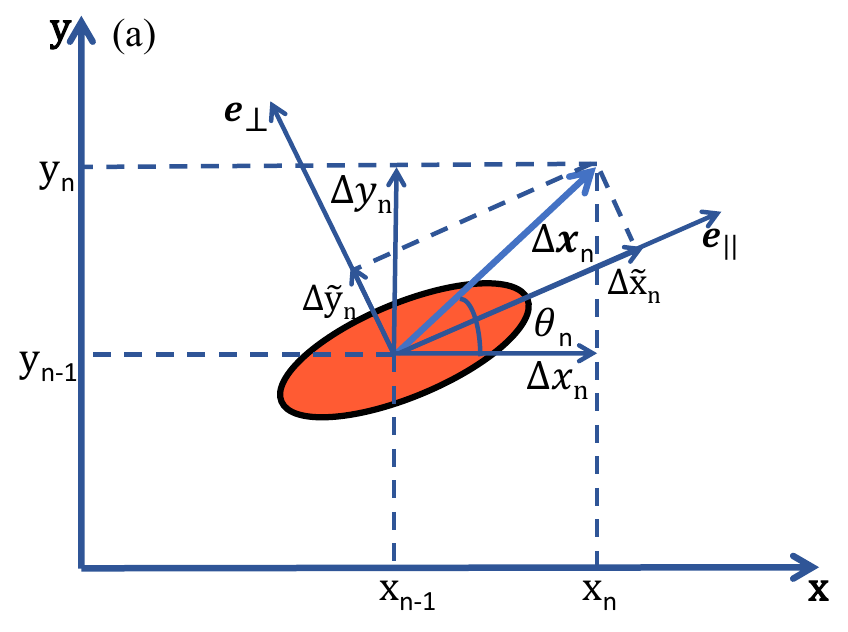}
    \includegraphics[width = \columnwidth]{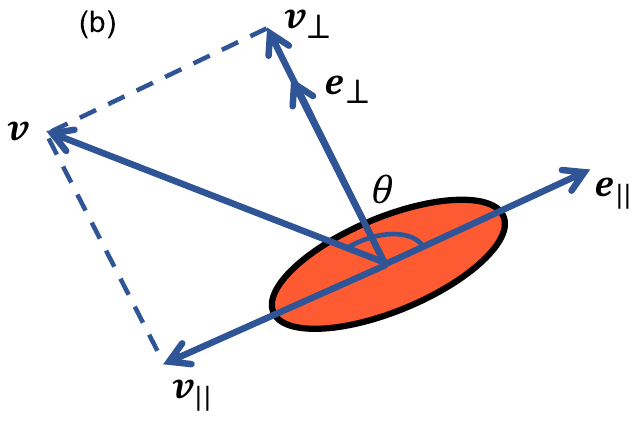}
    \caption{(a) Schematic representation of the transformation of the displacement vector $(\Delta x_n, \Delta y_n)$ from the laboratory frame to the body frame $(\Delta \tilde{x}_n, \Delta \tilde{y}_n)$ through the instantaneous orientation $\theta_n$ in the laboratory frame. (b) Decomposition of the velocity vector $\mathbf{v}$ along the major and minor axes of the tracer, defining a  velocity-body frame reference system $(\mathrm{v}_\parallel, \mathrm{v}_\perp)$.}
    \label{fig:frames}
\end{figure}

\renewcommand{\thefigure}{B\arabic{figure}}
\setcounter{figure}{0}
\begin{figure*}[t]
    \centering
    \includegraphics[width = 0.325\textwidth]{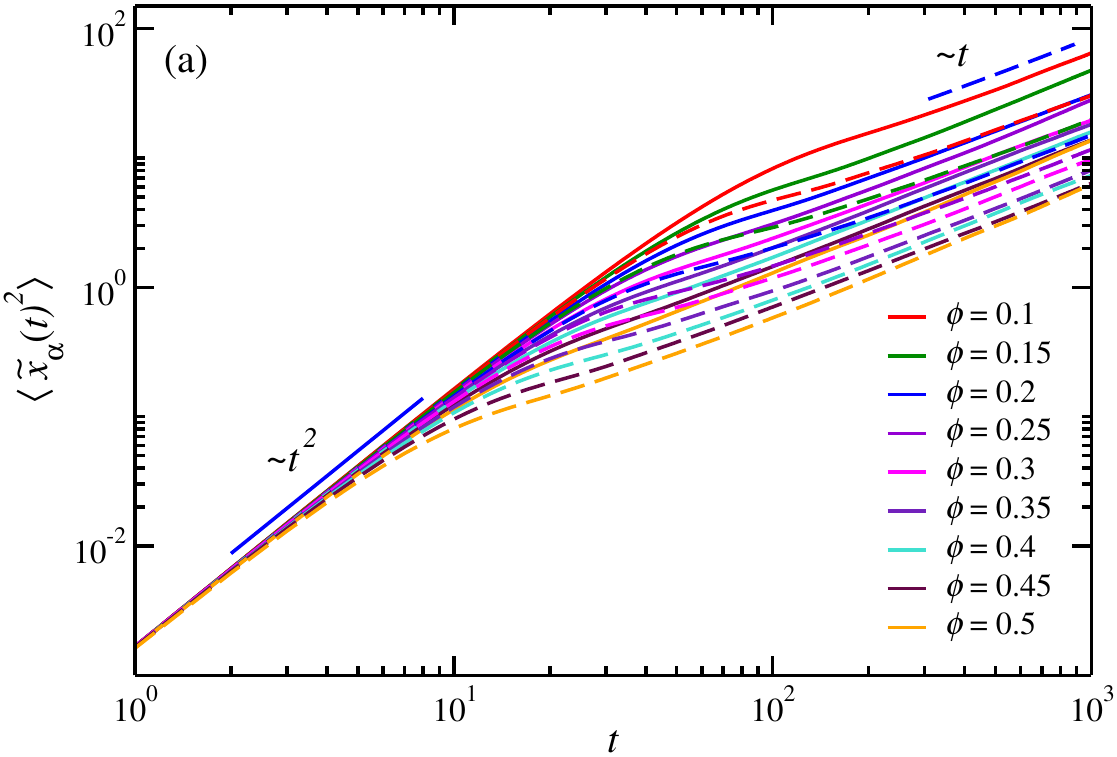}
    \includegraphics[width = 0.31\textwidth]{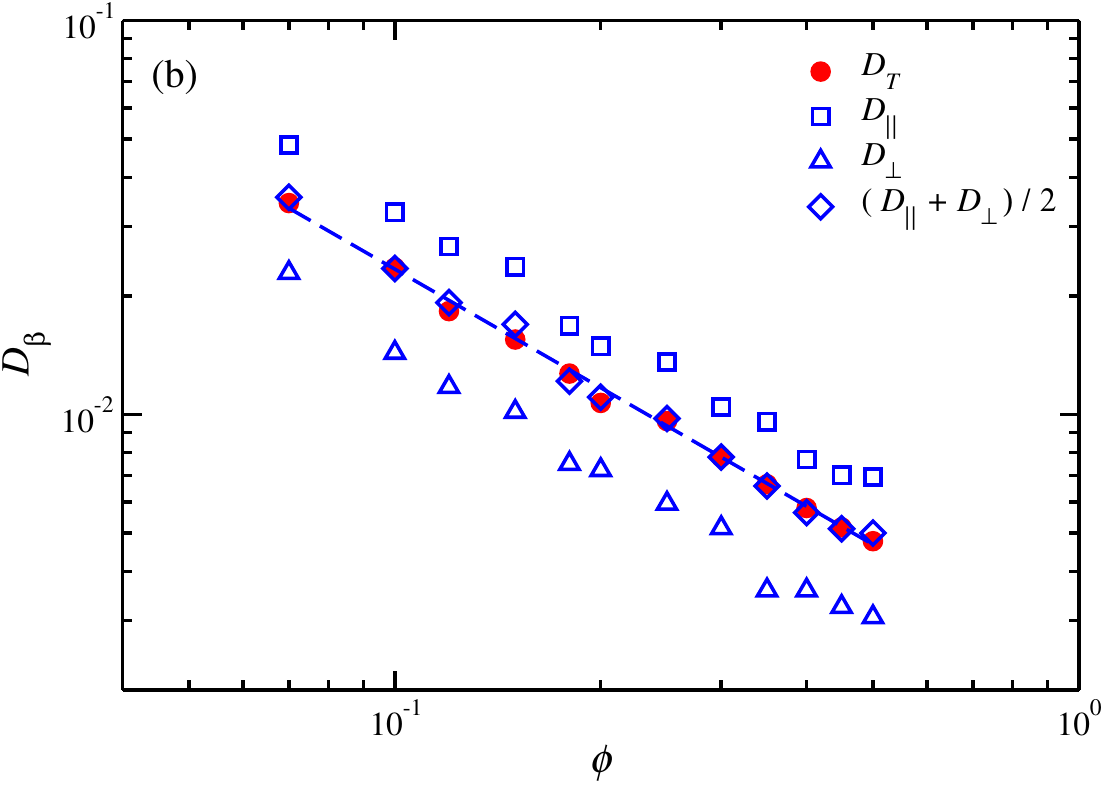}
    \includegraphics[width = 0.31\textwidth]{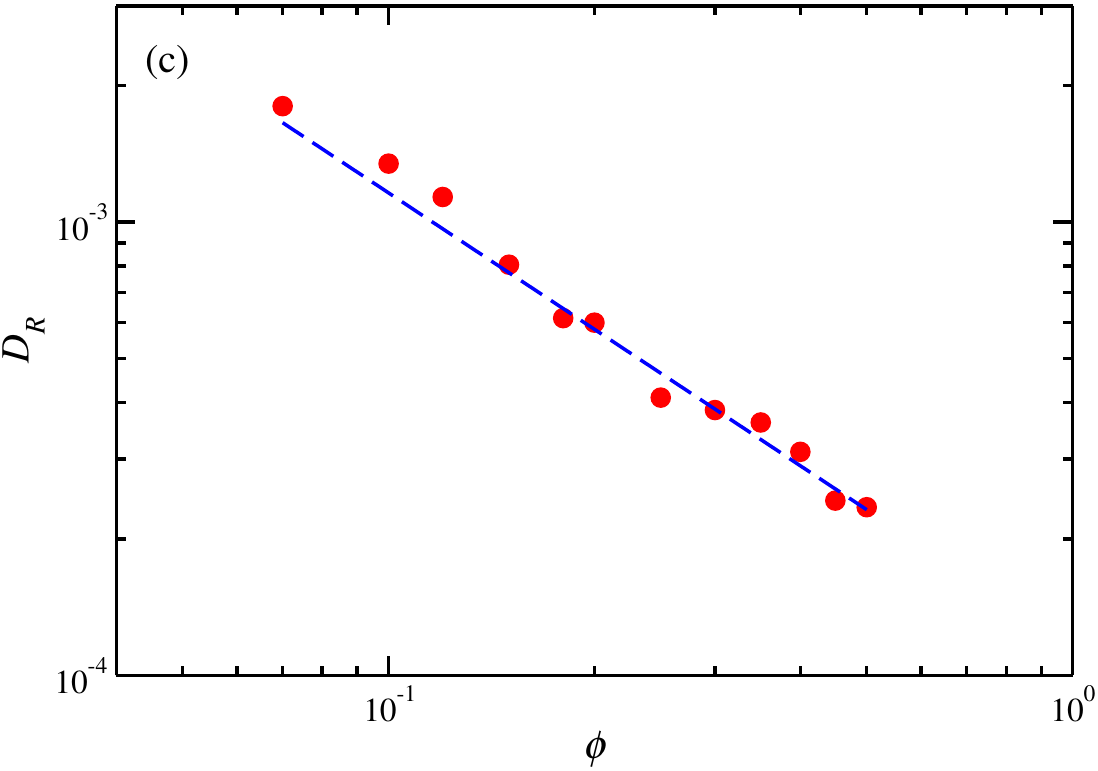}
    \caption{ (a) Mean-square displacements $\langle \tilde{x}_\alpha(t)^2\rangle$ of passive tracer along major and minor axis, $\alpha \in \{\parallel, \perp\}$, are shown by solid and dashed lines, respectively, for various area fractions $\phi$ of the passive dumbbells. The blue solid and dashed lines indicate ballistic and diffusive regimes, respectively. (b) Figure presents the center of mass diffusivity $D_T$ and the diffusivities along the major and minor axes ($D_\parallel$ and $D_\perp$) as a function of $\phi$. Diamond symbols present the relation $D_T = (D_\parallel + D_\perp)/2 $. (c) Rotational diffusivity $D_R$ is shown as a function of $\phi$. The blue dashed line in (b-c) displays the dependence $~\phi^{-1}$.}
    \label{fig:passive_results}
\end{figure*}

\subsection{Geometric body frame}
The geometric body frame is defined by the tracer’s principal axes, with unit vectors $\mathbf{e}_\parallel$ and $\mathbf{e}_\perp$ aligned along the major and minor axes, respectively. As shown in Figure~\ref{fig:frames}(a), this frame is rigidly attached to the center of the tracer and rotates with its orientation. To express the motion of tracer relative to this frame, the laboratory frame displacements $\Delta x_\alpha(t_n) = x_{\alpha }(t_n) - x_{\alpha}(t_{n-1})$ over the interval $\Delta t = t_n - t_{n-1}$ with $\alpha \in \{x, y\}$, is transformed into the geometric frame according to 
\begin{equation}
    \Delta \widetilde{x}_{\alpha}(t_n) = R_{\alpha\beta}(t_n) \Delta x_{\beta}(t_n).
    \label{eq:transform}
\end{equation}
Here $R_{\alpha \beta}$ is the rotation matrix    
\[
R_{\alpha \beta}(t_n)  =
\begin{pmatrix}
\cos\theta_n & \sin\theta_n \\
-\sin\theta_n &  \cos\theta_n
\end{pmatrix},
\] 
evaluated at the mid-step orientation $\theta_n = [\theta(t_n) + \theta(t_{n-1})]/2$, where $\theta(t)$ denotes the instantaneous angle of the major axis with respect to the $x$-axis of the laboratory frame. The total geometric body frame displacement is obtained by summing over displacements in each small time step $\widetilde{x}_\alpha(t) = \sum_{k=1}^{n} \Delta\widetilde{x}_\alpha(t_k)$, where $t_k = t_0 + t$ and $\alpha \in \{ \parallel, \perp \}$ denotes the displacement along the major and minor axes, respectively \cite{han2006brownian, peng2016diffusion}.

\subsection{Velocity-body frame}
To examine the tracer's motion relative to its instantaneous velocity direction, we introduce the velocity body frame. As illustrated in Figure~\ref{fig:frames}(b), this frame is constructed by projecting the tracer velocity $\mathbf{v}$ onto the geometric axes $\{\mathbf{e}_\parallel, \mathbf{e}_\perp\}$, yielding the components
$\mathrm{v}_\parallel = \mathbf{v} \cdot \mathbf{e}_\parallel$ and
$\mathrm{v}_\perp = \mathbf{v} \cdot \mathbf{e}_\perp$.
The corresponding unit vectors,
$\mathbf{e}_{\mathrm{v}_\parallel} = ({\mathrm{v_\parallel}}/{|\mathrm{v_\parallel}|})\mathbf{e}_\parallel $ and
$\mathbf{e}_{\mathrm{v}_\perp} = ({\mathrm{v_\perp}}/{|\mathrm{v_\perp}|}) \mathbf{e}_\perp$, define a coordinate system aligned with the instantaneous direction of motion. Any vector observable, such as a force $\mathbf{F}$ can then be decomposed in this frame as
$\mathrm{F}^\mathrm{v}_\parallel = \mathbf{F}\cdot\mathbf{e}_{\mathrm{v}_\parallel}$ and
$\mathrm{F}^\mathrm{v}_\perp = \mathbf{F}\cdot\mathbf{e}_{\mathrm{v}_\perp}$.

\section{Dynamics of elliptical tracer in equilibrium} 
\label{app:passive}

For comparison, we examine the equilibrium properties of an elliptical tracer in a bath of passive dumbbells, providing a reference for the tracer dynamics in the active bath. In the laboratory frame, the mean-square displacement (MSD) exhibits a short-time ballistic regime due to random thermal collisions with the dumbbells and crosses over at long times to a diffusive regime, $\langle\Delta \mathbf{R}^2(t)\rangle = 4D_T t$. Figure~\ref{fig:passive_results}(b) shows the translational diffusivity $D_T$ of the tracer as a function of the area fraction $\phi$ of passive dumbbells, which follows the dependence $D_T\sim\phi^{-1}$, consistent with the behavior of a circular passive tracer \cite{tiwari2024collective}.

The body-frame MSD is obtained by transforming the laboratory-frame displacements (Appendix~\ref{app:frames}). The axis-resolved MSDs, $\langle \widetilde{x}_\alpha(t)^2\rangle$ with $\alpha\in\{\parallel,\perp\}$, display the same qualitative behavior as the center-of-mass MSD: super-diffusive at short times and diffusive at long times, $\langle \widetilde{x}_\alpha(t)^2\rangle = 2D_\alpha t$, as shown in Figure~\ref{fig:passive_results}(a). The corresponding diffusivities $D_\parallel$ and $D_\perp$ decrease with  $\phi^{-1}$, see Figure~\ref{fig:passive_results}(b), with $D_\parallel > D_\perp$. Moreover, the relation $D_T = (D_\parallel + D_\perp)/2 $ holds, in agreement with previous observations \cite{han2006brownian}.

The rotational diffusivity is obtained from the exponential decay of the orientation correlation function of the major axis of the tracer, $C_\theta(t)=\langle \cos[\theta(t) -\theta(0)] \rangle\sim \exp(-t/\tau_r)$, where $\tau_r = 1/D_R$. Similar to $D_T$, the rotational diffusivity $D_R$ decreases approximately as $\phi^{-1}$, as illustrated in Figure~\ref{fig:passive_results}(c).

\section{Variance of active force on tracer} 
\label{app:diffusivity}
The speed and diffusivity can be expressed in terms of the variance of the effective force from the surrounding medium  (active dumbbells) given as
\begin{equation}
    \langle F_e^2\rangle =  \left \langle \left( -\sum_{i=1}^{N_c}\mathbf{F}^i_m\right)^2\right \rangle,    
\end{equation}
where $N_c$ is the number of monomers interacting with the tracer. Further, $\langle F_e^2\rangle$ expands into self-terms and cross-terms
\begin{equation}
    \langle F_e^2\rangle = \left \langle \sum_{i=1}^{N_c}  (\mathbf{F}^i_m )^2 \right \rangle + \left \langle \sum_{i = 1}^{N_c} \psum_{j= 1}^{N_c}\mathbf{F}^i_m \cdot \mathbf{F}^j_m \right \rangle.
\end{equation}
Here, the prime in the summation excludes the $i=j$ terms. Denoting $\sigma^2_F = \langle (\mathbf{F}^i_m)^2 \rangle$ as the self-force variance of a single monomer interacting with the tracer, and $c_\mathrm{F} = \langle \mathbf{F}^i_m \cdot \mathbf{F}^j_m\rangle$ as the cross-force correlation between the $i^{\mathrm{th}}$ and $j^{\mathrm{th}}$ monomers in the vicinity of the tracer, with $\langle N_c \rangle$ the mean number of monomers interacting with the tracer, we obtain
\begin{equation}
     \langle F_e^2\rangle = \langle N_c \rangle  \sigma^2_\mathrm{F} +  \langle N_c(N_c - 1) \rangle c_F.
     \label{eq:var_fsq}
\end{equation}

The decomposition of the total force variance [Eq.~\ref{eq:var_fsq}] into the self-force contribution, $\sigma^2_\mathrm{F}$, and the cross-correlation contribution, $c_\mathrm{F}$, is estimated numerically from our simulations. Figure~\ref{fig:cross_sq_appendix} shows that $\sigma^2_\mathrm{F}$ increases monotonically with $\phi$, reflecting more frequent collisions as the bath becomes denser. In contrast, $c_\mathrm{F}$ decreases at larger $\phi$, indicating increasing structural organization of the active dumbbells around the tracer.

The dumbbells accumulate and align along the tracer surface, with each dumbbell exerting a force on the tracer from its local contact point. Consequently, the forces exerted at different locations along the tracer periphery become less correlated. Similar behavior has been reported for active dumbbells interacting with a circular static obstacle in Ref.~\onlinecite{tiwari2024collective}. This arrangement becomes more pronounced as crowding increases, leading to a decrease in the cross-correlation between forces exerted by different dumbbells.



\renewcommand{\thefigure}{C\arabic{figure}}
\setcounter{figure}{0}
\begin{figure}[t]
    \centering
    \includegraphics[width = \columnwidth]{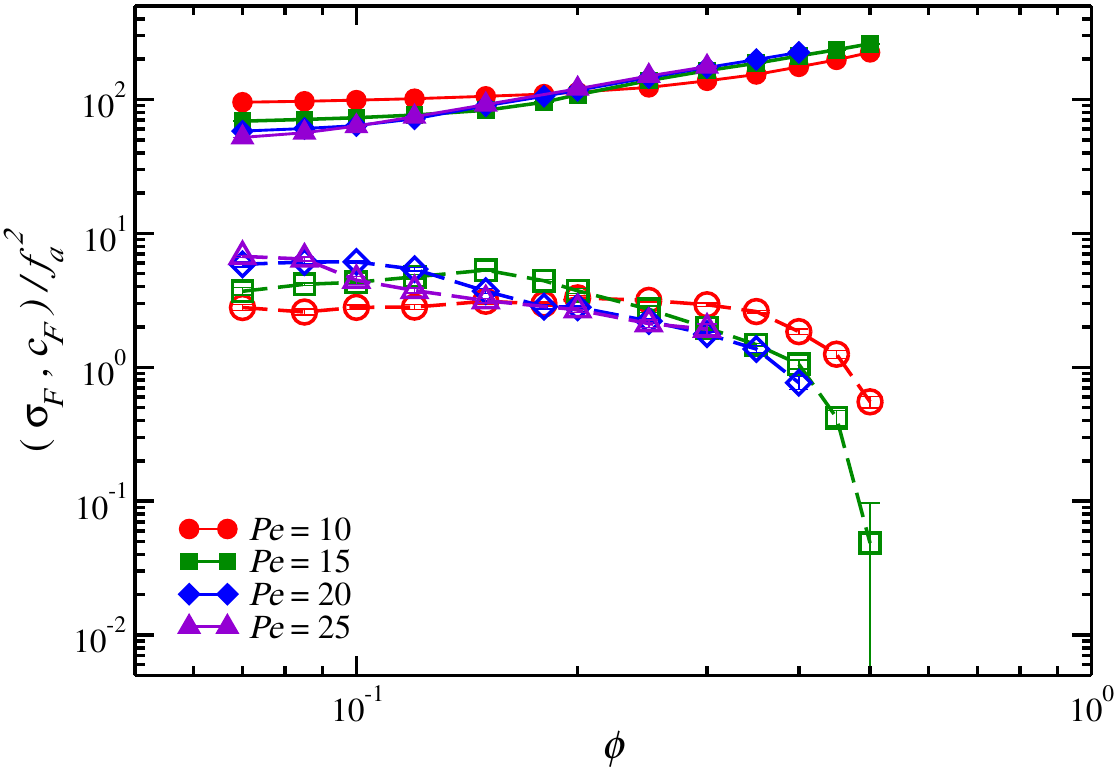}
    \caption{ Self-force variance of force of a monomer $\sigma_F^2$ interacting with the tracer (solid symbols), and cross-force correlation $c_F$ (open symbols), as a function of area fraction $\phi$ for various $Pe$. }
    \label{fig:cross_sq_appendix}
\end{figure}


\newpage

 \bibliography{reference}

@article{ermak1978brownian,
  title={Brownian dynamics with hydrodynamic interactions},
  author={Ermak, Donald L and McCammon, J Andrew},
  journal={The Journal of chemical physics},
  volume={69},
  number={4},
  pages={1352--1360},
  year={1978},
  publisher={American Institute of Physics}
}

@article{Spotswood_D_Stoddard,
  title={Identifying Clusters in Computer Experiments on Systems of Particles},
  author={Spotswood D Stoddard},
  journal={Journal of Computational Physics},
  volume={27},
  number={2},
  pages={291--293},
  year={1978}
}

@article{marchetti2013hydrodynamics,
  title={Hydrodynamics of soft active matter},
  author={Marchetti, M Cristina and Joanny, Jean-Fran{\c{c}}ois and Ramaswamy, Sriram and Liverpool, Tanniemola B and Prost, Jacques and Rao, Madan and Simha, R Aditi},
  journal={Reviews of modern physics},
  volume={85},
  number={3},
  pages={1143},
  year={2013},
  publisher={APS}
}

@article{cates2015motility,
  title={Motility-induced phase separation},
  author={Cates, Michael E and Tailleur, Julien},
  journal={Annu. Rev. Condens. Matter Phys.},
  volume={6},
  number={1},
  pages={219--244},
  year={2015},
  publisher={Annual Reviews}
}

@article{suma2014motility,
  title={Motility-induced phase separation in an active dumbbell fluid},
  author={Suma, Antonio and Gonnella, Giuseppe and Marenduzzo, Davide and Orlandini, Enzo},
  journal={Europhysics Letters},
  volume={108},
  number={5},
  pages={56004},
  year={2014},
  publisher={IOP Publishing}
}

@article{suma2014dynamics,
  title={Dynamics of a homogeneous active dumbbell system},
  author={Suma, Antonio and Gonnella, Giuseppe and Laghezza, Gianluca and Lamura, Antonio and Mossa, Alessandro and Cugliandolo, Leticia F},
  journal={Physical Review E},
  volume={90},
  number={5},
  pages={052130},
  year={2014},
  publisher={APS}
}

@article{drescher2011fluid,
  title={Fluid dynamics and noise in bacterial cell--cell and cell--surface scattering},
  author={Drescher, Knut and Dunkel, J{\"o}rn and Cisneros, Luis H and Ganguly, Sujoy and Goldstein, Raymond E},
  journal={Proceedings of the National Academy of Sciences},
  volume={108},
  number={27},
  pages={10940--10945},
  year={2011},
  publisher={National Acad Sciences}
}

@article{wu2000particle,
  title={Particle diffusion in a quasi-two-dimensional bacterial bath},
  author={Wu, Xiao-Lun and Libchaber, Albert},
  journal={Physical review letters},
  volume={84},
  number={13},
  pages={3017},
  year={2000},
  publisher={APS}
}

@article{mallory2014curvature,
  title={Curvature-induced activation of a passive tracer in an active bath},
  author={Mallory, SA and Valeriani, C and Cacciuto, A},
  journal={Physical Review E},
  volume={90},
  number={3},
  pages={032309},
  year={2014},
  publisher={APS}
}

@book{Doi2013SoftMP,
    author = {Doi, Masao},
    isbn = {9780199652952},
    title = "{Liquid crystals}",
    booktitle = "{Soft Matter Physics}",
    publisher = {Oxford University Press},
    year = {2013},
    month = {06},}

@book{allen2017computer,
  title={Computer simulation of liquids},
  author={Allen, Michael P and Tildesley, Dominic J},
  year={2017},
  publisher={Oxford university press}
}

@article{martin2018active,
  title={Active brownian filamentous polymers under shear flow},
  author={Mart{\'\i}n-G{\'o}mez, Aitor and Gompper, Gerhard and Winkler, Roland},
  journal={Polymers},
  volume={10},
  number={8},
  pages={837},
  year={2018},
  publisher={Multidisciplinary Digital Publishing Institute}
}

@article{angelani2009self,
  title={Self-starting micromotors in a bacterial bath},
  author={Angelani, Luca and Di Leonardo, Roberto and Ruocco, Giancarlo},
  journal={Physical review letters},
  volume={102},
  number={4},
  pages={048104},
  year={2009},
  publisher={APS}
}

@article{sokolov2010swimming,
  title={Swimming bacteria power microscopic gears},
  author={Sokolov, Andrey and Apodaca, Mario M and Grzybowski, Bartosz A and Aranson, Igor S},
  journal={Proceedings of the National Academy of Sciences},
  volume={107},
  number={3},
  pages={969--974},
  year={2010},
  publisher={National Acad Sciences}
}

@article{knevzevic2020effective,
  title={Effective Langevin equations for a polar tracer in an active bath},
  author={Kne{\v{z}}evi{\'c}, Milo{\v{s}} and Stark, Holger},
  journal={New Journal of Physics},
  volume={22},
  number={11},
  pages={113025},
  year={2020},
  publisher={IOP Publishing}
}

@article{jose2021phase,
  title={Phase separation of an active colloidal suspension via quorum-sensing},
  author={Jose, Francis and Anand, Shalabh K and Singh, Sunil P},
  journal={Soft Matter},
  volume={17},
  number={11},
  pages={3153--3161},
  year={2021},
  publisher={Royal Society of Chemistry}
}

@article{mino2011enhanced,
  title={Enhanced diffusion due to active swimmers at a solid surface},
  author={Mino, Gast{\'o}n and Mallouk, Thomas E and Darnige, Thierry and Hoyos, Mauricio and Dauchet, Jeremi and Dunstan, Jocelyn and Soto, Rodrigo and Wang, Yang and Rousselet, Annie and Clement, Eric},
  journal={Physical review letters},
  volume={106},
  number={4},
  pages={048102},
  year={2011},
  publisher={APS}
}

@article{jepson2013enhanced,
  title={Enhanced diffusion of nonswimmers in a three-dimensional bath of motile bacteria},
  author={Jepson, Alys and Martinez, Vincent A and Schwarz-Linek, Jana and Morozov, Alexander and Poon, Wilson CK},
  journal={Physical Review E},
  volume={88},
  number={4},
  pages={041002},
  year={2013},
  publisher={APS}
}

@article{patteson2016particle,
  title={Particle diffusion in active fluids is non-monotonic in size},
  author={Patteson, Alison E and Gopinath, Arvind and Purohit, Prashant K and Arratia, Paulo E},
  journal={Soft matter},
  volume={12},
  number={8},
  pages={2365--2372},
  year={2016},
  publisher={Royal Society of Chemistry}
}

@article{tiwari2024collective,
  title={Collective dynamics of active dumbbells near a circular obstacle},
  author={Tiwari, Chandranshu and Singh, Sunil P},
  journal={Soft Matter},
  volume={20},
  number={24},
  pages={4816--4826},
  year={2024},
  publisher={Royal Society of Chemistry}
}

@article{peng2016diffusion,
  title={Diffusion of ellipsoids in bacterial suspensions},
  author={Peng, Yi and Lai, Lipeng and Tai, Yi-Shu and Zhang, Kechun and Xu, Xinliang and Cheng, Xiang},
  journal={Physical review letters},
  volume={116},
  number={6},
  pages={068303},
  year={2016},
  publisher={APS}
}

@article{yang2016dynamics,
  title={Dynamics of ellipsoidal tracers in swimming algal suspensions},
  author={Yang, Ou and Peng, Yi and Liu, Zhengyang and Tang, Chao and Xu, Xinliang and Cheng, Xiang},
  journal={Physical Review E},
  volume={94},
  number={4},
  pages={042601},
  year={2016},
  publisher={APS}
}

@article{li2024anisotropic,
  title={Anisotropic diffusion of elongated particles in active coherent flows},
  author={Li, Dongdong and Liu, Yanan and Luo, Hao and Jing, Guangyin},
  journal={Micromachines},
  volume={15},
  number={2},
  pages={199},
  year={2024},
  publisher={MDPI}
}

@article{brangwynne2008cytoplasmic,
  title={Cytoplasmic diffusion: molecular motors mix it up},
  author={Brangwynne, Clifford P and Koenderink, Gijsje H and MacKintosh, Frederick C and Weitz, David A},
  journal={The Journal of cell biology},
  volume={183},
  number={4},
  pages={583--587},
  year={2008},
  publisher={Rockefeller University Press}
}

@article{koslover2017cytoplasmic,
  title={Cytoplasmic flow and mixing due to deformation of motile cells},
  author={Koslover, Elena F and Chan, Caleb K and Theriot, Julie A},
  journal={Biophysical journal},
  volume={113},
  number={9},
  pages={2077--2087},
  year={2017},
  publisher={Elsevier}
}

@article{lu2023go,
  title={Go with the flow--bulk transport by molecular motors},
  author={Lu, Wen and Gelfand, Vladimir I},
  journal={Journal of cell science},
  volume={136},
  number={5},
  pages={jcs260300},
  year={2023},
  publisher={The Company of Biologists Ltd}
}

@article{ziepke2022multi,
  title={Multi-scale organization in communicating active matter},
  author={Ziepke, Alexander and Maryshev, Ivan and Aranson, Igor S and Frey, Erwin},
  journal={Nature communications},
  volume={13},
  number={1},
  pages={6727},
  year={2022},
  publisher={Nature Publishing Group UK London}
}

@article{lin2025interactions,
  title={Interactions Between Active Matters and Endogenous Fields},
  author={Lin, Jinwei and Guan, Qiaoxin and Feng, Jiangqi and Chen, Shuqin and Xu, Leilei and Guan, Jianguo and S{\'a}nchez, Samuel},
  journal={Advanced Materials},
  pages={e03091},
  year={2025},
  publisher={Wiley Online Library}
}

@article{kapellos2022impact,
  title={Impact of microbial uptake on the nutrient plume around marine organic particles: high-resolution numerical analysis},
  author={Kapellos, George E and Eberl, Hermann J and Kalogerakis, Nicolas and Doyle, Patrick S and Paraskeva, Christakis A},
  journal={Microorganisms},
  volume={10},
  number={10},
  pages={2020},
  year={2022},
  publisher={MDPI}
}

@article{mogre2020getting,
  title={Getting around the cell: physical transport in the intracellular world},
  author={Mogre, Saurabh S and Brown, Aidan I and Koslover, Elena F},
  journal={Physical Biology},
  volume={17},
  number={6},
  pages={061003},
  year={2020},
  publisher={IOP Publishing}
}

@article{morozov2014enhanced,
  title={Enhanced diffusion of tracer particles in dilute bacterial suspensions},
  author={Morozov, Alexander and Marenduzzo, Davide},
  journal={Soft Matter},
  volume={10},
  number={16},
  pages={2748--2758},
  year={2014},
  publisher={Royal Society of Chemistry}
}

@article{kurtuldu2011enhancement,
  title={Enhancement of biomixing by swimming algal cells in two-dimensional films},
  author={Kurtuldu, H{\"u}seyin and Guasto, Jeffrey S and Johnson, Karl A and Gollub, Jerry P},
  journal={Proceedings of the National Academy of Sciences},
  volume={108},
  number={26},
  pages={10391--10395},
  year={2011},
  publisher={National Academy of Sciences}
}

@article{granek2022anomalous,
  title={Anomalous transport of tracers in active baths},
  author={Granek, Omer and Kafri, Yariv and Tailleur, Julien},
  journal={Physical Review Letters},
  volume={129},
  number={3},
  pages={038001},
  year={2022},
  publisher={APS}
}

@article{maggi2017memory,
  title={Memory-less response and violation of the fluctuation-dissipation theorem in colloids suspended in an active bath},
  author={Maggi, Claudio and Paoluzzi, Matteo and Angelani, Luca and Di Leonardo, Roberto},
  journal={Scientific reports},
  volume={7},
  number={1},
  pages={17588},
  year={2017},
  publisher={Nature Publishing Group UK London}
}

@article{grossmann2024non,
  title={Non-Gaussian displacements in active transport on a carpet of motile cells},
  author={Gro{\ss}mann, Robert and Bort, Lara S and Moldenhawer, Ted and Stange, Maike and Panah, Setareh Sharifi and Metzler, Ralf and Beta, Carsten},
  journal={Physical Review Letters},
  volume={132},
  number={8},
  pages={088301},
  year={2024},
  publisher={APS}
}

@article{von2021diffusive,
  title={Diffusive dynamics of elongated particles in active colloidal suspensions of motile algae},
  author={von R{\"u}ling, Florian and Kolley, Francine and Eremin, Alexey},
  journal={Colloid and Polymer Science},
  volume={299},
  number={2},
  pages={289--296},
  year={2021},
  publisher={Springer}
}

@article{chen2007fluctuations,
  title={Fluctuations and rheology in active bacterial suspensions},
  author={Chen, Daniel TN and Lau, AWC and Hough, Lawrence A and Islam, Mohammad F and Goulian, Mark and Lubensky, Thomas C and Yodh, Arjun G},
  journal={Physical review letters},
  volume={99},
  number={14},
  pages={148302},
  year={2007},
  publisher={APS}
}

@article{seyforth2022nonequilibrium,
  title={Nonequilibrium fluctuations and nonlinear response of an active bath},
  author={Seyforth, Hunter and Gomez, Mauricio and Rogers, W Benjamin and Ross, Jennifer L and Ahmed, Wylie W},
  journal={Physical review research},
  volume={4},
  number={2},
  pages={023043},
  year={2022},
  publisher={APS}
}

@article{nordanger2022anisotropic,
  title={Anisotropic diffusion of ellipsoidal tracers in microswimmer suspensions},
  author={Nordanger, Henrik and Morozov, Alexander and Stenhammar, Joakim},
  journal={Physical Review Fluids},
  volume={7},
  number={1},
  pages={013103},
  year={2022},
  publisher={APS}
}

@article{han2006brownian,
  title={Brownian motion of an ellipsoid},
  author={Han, Yilong and Alsayed, Ahmed M and Nobili, Maurizio and Zhang, Jian and Lubensky, Tom C and Yodh, Arjun G},
  journal={Science},
  volume={314},
  number={5799},
  pages={626--630},
  year={2006},
  publisher={American Association for the Advancement of Science}
}

@article{maggi2016self,
  title={Self-Assembly of Micromachining Systems Powered by Janus Micromotors},
  author={Maggi, Claudio and Simmchen, Juliane and Saglimbeni, Filippo and Katuri, Jaideep and Dipalo, Michele and De Angelis, Francesco and Sanchez, Samuel and Di Leonardo, Roberto and others},
  journal={SMALL},
  volume={12},
  number={4},
  pages={446--451},
  year={2016},
  publisher={Wiley-VCH Verlag}
}

@article{codina2022small,
  title={Small obstacle in a large polar flock},
  author={Codina, Joan and Mahault, Beno{\^\i}t and Chat{\'e}, Hugues and Dobnikar, Jure and Pagonabarraga, Ignacio and Shi, Xia-qing},
  journal={Physical Review Letters},
  volume={128},
  number={21},
  pages={218001},
  year={2022},
  publisher={APS}
}

@article{pietzonka2019autonomous,
  title={Autonomous engines driven by active matter: Energetics and design principles},
  author={Pietzonka, Patrick and Fodor, {\'E}tienne and Lohrmann, Christoph and Cates, Michael E and Seifert, Udo},
  journal={Physical Review X},
  volume={9},
  number={4},
  pages={041032},
  year={2019},
  publisher={APS}
}

@article{ underhill2008diffusion,
  title={Diffusion and spatial correlations in suspensions of swimming particles},
  author={Underhill, Patrick T and Hernandez-Ortiz, Juan P and Graham, Michael D},
  journal={Physical review letters},
  volume={100},
  number={24},
  pages={248101},
  year={2008},
  publisher={APS}
}

@article{caprini2024emergent,
  title={Emergent memory from tapping collisions in active granular matter},
  author={Caprini, Lorenzo and Ldov, Anton and Gupta, Rahul Kumar and Ellenberg, Hendrik and Wittmann, Ren{\'e} and L{\"o}wen, Hartmut and Scholz, Christian},
  journal={Communications Physics},
  volume={7},
  number={1},
  pages={52},
  year={2024},
  publisher={Nature Publishing Group UK London}
}

@article{dhar2024active,
  title={Active transport of a passive colloid in a bath of run-and-tumble particles},
  author={Dhar, Tanumoy and Saintillan, David},
  journal={Scientific Reports},
  volume={14},
  number={1},
  pages={11844},
  year={2024},
  publisher={Nature Publishing Group UK London}
}

@article{kasyap2014hydrodynamic,
  title={Hydrodynamic tracer diffusion in suspensions of swimming bacteria},
  author={Kasyap, TV and Koch, Donald L and Wu, Mingming},
  journal={Physics of Fluids},
  volume={26},
  number={8},
  year={2014},
  publisher={AIP Publishing}
}

@article{burkholder2017tracer,
  title={Tracer diffusion in active suspensions},
  author={Burkholder, Eric W and Brady, John F},
  journal={Physical Review E},
  volume={95},
  number={5},
  pages={052605},
  year={2017},
  publisher={APS}
}

@article{argun2016non,
  title={Non-Boltzmann stationary distributions and nonequilibrium relations in active baths},
  author={Argun, Aykut and Moradi, Ali-Reza and Pin{\c{c}}e, Er{\c{c}}aǧ and Bagci, Gokhan Baris and Imparato, Alberto and Volpe, Giovanni},
  journal={Physical Review E},
  volume={94},
  number={6},
  pages={062150},
  year={2016},
  publisher={APS}
}

@article{yadav2026behavior,
  title={Behavior of passive polymeric tracers of different topologies in a dilute bath of active Brownian particles},
  author={Yadav, Ramanand Singh and Metzler, Ralf and Chakrabarti, Rajarshi},
  journal={Physical Review Research},
  volume={8},
  number={1},
  pages={013053},
  year={2026},
  publisher={APS}
}

@article{zhao2017enhanced,
  title={Enhanced diffusion of passive tracers in active enzyme solutions},
  author={Zhao, Xi and Dey, Krishna K and Jeganathan, Selva and Butler, Peter J and C{\'o}rdova-Figueroa, Ubaldo M and Sen, Ayusman},
  journal={Nano letters},
  volume={17},
  number={8},
  pages={4807--4812},
  year={2017},
  publisher={ACS Publications}
}

@article{ortlieb2019statistics,
  title={Statistics of colloidal suspensions stirred by microswimmers},
  author={Ortlieb, Levke and Rafa{\"\i}, Salima and Peyla, Philippe and Wagner, Christian and John, Thomas},
  journal={Physical review letters},
  volume={122},
  number={14},
  pages={148101},
  year={2019},
  publisher={APS}
}

@article{fodor2015activity,
  title={Activity-driven fluctuations in living cells},
  author={Fodor, {\'E} and Guo, M and Gov, NS and Visco, P and Weitz, DA and Van Wijland, F},
  journal={Europhysics Letters},
  volume={110},
  number={4},
  pages={48005},
  year={2015},
  publisher={EDP Sciences, IOP Publishing and Societ{\`a} Italiana di Fisica}
}

@article{otten2012local,
  title={Local motion analysis reveals impact of the dynamic cytoskeleton on intracellular subdiffusion},
  author={Otten, Marcus and Nandi, Amitabha and Arcizet, Delphine and Gorelashvili, Mari and Lindner, Benjamin and Heinrich, Doris},
  journal={Biophysical journal},
  volume={102},
  number={4},
  pages={758--767},
  year={2012},
  publisher={Elsevier}
}

@article{souza2017anomalous,
  title={Anomalous diffusion and q-Weibull velocity distributions in epithelial cell migration},
  author={Souza Vilela Podest{\'a}, Tatiane and Venzel Rosembach, Tiago and Aparecida dos Santos, An{\'e}sia and Lobato Martins, Marcelo},
  journal={Plos one},
  volume={12},
  number={7},
  pages={e0180777},
  year={2017},
  publisher={Public Library of Science San Francisco, CA USA}
}

@article{dieterich2008anomalous,
  title={Anomalous dynamics of cell migration},
  author={Dieterich, Peter and Klages, Rainer and Preuss, Roland and Schwab, Albrecht},
  journal={Proceedings of the National Academy of Sciences},
  volume={105},
  number={2},
  pages={459--463},
  year={2008},
  publisher={National Academy of Sciences}
}

@article{ye2020active,
  title={Active noise experienced by a passive particle trapped in an active bath},
  author={Ye, Simin and Liu, Peng and Ye, Fangfu and Chen, Ke and Yang, Mingcheng},
  journal={Soft matter},
  volume={16},
  number={19},
  pages={4655--4660},
  year={2020},
  publisher={Royal Society of Chemistry}
}

@article{villalobos2025active,
  title={Active bacterial baths in droplets},
  author={Villalobos-Concha, Cristian and Liu, Zhengyang and Ramos, Gabriel and Goral, Martyna and Lindner, Anke and L{\'o}pez-Le{\'o}n, Teresa and Cl{\'e}ment, Eric and Soto, Rodrigo and Cordero, Mar{\'\i}a Luisa},
  journal={Proceedings of the National Academy of Sciences},
  volume={122},
  number={31},
  pages={e2426096122},
  year={2025},
  publisher={National Academy of Sciences}
}

@article{feng2023unraveling,
  title={Unraveling on kinesin acceleration in intracellular environments: A theory for active bath},
  author={Feng, Mengkai and Hou, Zhonghuai},
  journal={Physical Review Research},
  volume={5},
  number={1},
  pages={013206},
  year={2023},
  publisher={APS}
}

@article{baule2023universal,
  title={Universal Poisson statistics of a passive tracer diffusing in dilute active suspensions},
  author={Baule, Adrian},
  journal={Proceedings of the National Academy of Sciences},
  volume={120},
  number={50},
  pages={e2308226120},
  year={2023},
  publisher={National Academy of Sciences}
}

@article{aporvari2020anisotropic,
  title={Anisotropic dynamics of a self-assembled colloidal chain in an active bath},
  author={Aporvari, Mehdi Shafiei and Utkur, Mustafa and Saritas, Emine Ulku and Volpe, Giovanni and Stenhammar, Joakim},
  journal={Soft Matter},
  volume={16},
  number={24},
  pages={5609--5614},
  year={2020},
  publisher={Royal Society of Chemistry}
}

@article{kaiser2014transport,
  title={Transport powered by bacterial turbulence},
  author={Kaiser, Andreas and Peshkov, Anton and Sokolov, Andrey and Ten Hagen, Borge and L{\"o}wen, Hartmut and Aranson, Igor S},
  journal={Physical review letters},
  volume={112},
  number={15},
  pages={158101},
  year={2014},
  publisher={APS}
}

@article{pellicciotta2025wall,
  title={Wall torque controls propulsion of curved microstructures in bacterial baths},
  author={Pellicciotta, Nicola and Bagal, Ojus Satish and Cannarsa, Maria Cristina and Bianchi, Silvio and Di Leonardo, Roberto},
  journal={Physical Review Letters},
  volume={135},
  number={13},
  pages={138302},
  year={2025},
  publisher={APS}
}

@article{semeraro2018effective,
  title={Effective interactions and dynamics of small passive particles in an active bacterial medium},
  author={Semeraro, Enrico F and Devos, Juliette M and Narayanan, Theyencheri},
  journal={The Journal of Chemical Physics},
  volume={148},
  number={20},
  year={2018},
  publisher={AIP Publishing}
}

@article{lagarde2020colloidal,
  title={Colloidal transport in bacteria suspensions: From bacteria collision to anomalous and enhanced diffusion},
  author={Lagarde, Antoine and Dag{\`e}s, No{\'e}mie and Nemoto, Takahiro and D{\'e}mery, Vincent and Bartolo, Denis and Gibaud, Thomas},
  journal={Soft Matter},
  volume={16},
  number={32},
  pages={7503--7512},
  year={2020},
  publisher={Royal Society of Chemistry}
}

@article{paul2026interplay,
  title={Interplay of activity and non-reciprocity in tracer dynamics: From non-equilibrium fluctuation-dissipation to giant diffusion},
  author={Paul, Subhajit and Chaudhuri, Debasish},
  journal={arXiv preprint arXiv:2601.03591},
  year={2026}
}

@article{daza2025diffusion,
  title={Diffusion of tracer particles in early growing biofilms a computer simulation study},
  author={Daza, Fabi{\'a}n A Garc{\'\i}a and Rodr{\'\i}guez-Rivas, {\'A}lvaro and Govantes, Fernando and Cuetos, Alejandro},
  journal={Colloids and Surfaces B: Biointerfaces},
  volume={255},
  pages={114903},
  year={2025},
  publisher={Elsevier}
}

@article{martin2021statistical,
  title={Statistical mechanics of active Ornstein-Uhlenbeck particles},
  author={Martin, David and O'Byrne, J{\'e}r{\'e}my and Cates, Michael E and Fodor, {\'E}tienne and Nardini, Cesare and Tailleur, Julien and Van Wijland, Fr{\'e}d{\'e}ric},
  journal={Physical Review E},
  volume={103},
  number={3},
  pages={032607},
  year={2021},
  publisher={APS}
}

@article{szamel2014self,
  title={Self-propelled particle in an external potential: Existence of an effective temperature},
  author={Szamel, Grzegorz},
  journal={Physical Review E},
  volume={90},
  number={1},
  pages={012111},
  year={2014},
  publisher={APS}
}

@book{zwanzig2001nonequilibrium,
  title={Nonequilibrium statistical mechanics},
  author={Zwanzig, Robert},
  year={2001},
  publisher={Oxford university press}
}

@article{maggi2014generalized,
  title={Generalized energy equipartition in harmonic oscillators driven by active baths},
  author={Maggi, Claudio and Paoluzzi, Matteo and Pellicciotta, Nicola and Lepore, Alessia and Angelani, Luca and Di Leonardo, Roberto},
  journal={Physical review letters},
  volume={113},
  number={23},
  pages={238303},
  year={2014},
  publisher={APS}
}

@article{nikola2016active,
  title={Active particles with soft and curved walls: Equation of state, ratchets, and instabilities},
  author={Nikola, Nikolai and Solon, Alexandre P and Kafri, Yariv and Kardar, Mehran and Tailleur, Julien and Voituriez, Rapha{\"e}l},
  journal={Physical review letters},
  volume={117},
  number={9},
  pages={098001},
  year={2016},
  publisher={APS}
}

\end{document}